\documentclass[10pt,journal,compsoc]{IEEEtran}
\ifCLASSOPTIONcompsoc
  \usepackage[nocompress]{cite}
\else
  \usepackage{cite}
\fi
\ifCLASSINFOpdf
  \usepackage[pdftex]{graphicx}
  \graphicspath{{images/}}
\else
\fi
\usepackage{amsmath}
\usepackage[noend]{algpseudocode}
\usepackage{algorithm}

\usepackage{array}

\usepackage{caption}
\usepackage{subcaption}
\usepackage{url}

\usepackage{amssymb}
\usepackage{epsfig}
\usepackage{enumerate}
\usepackage{amsthm}
\usepackage{multirow}
\usepackage{color}
\usepackage{xcolor}

\theoremstyle{definition}
\newtheorem{dfn}{Definition}[section]

\theoremstyle{plain}
\newtheorem{thm}{Theorem}[section]
\newtheorem{lem}{Lemma}[section]

\newtheorem*{thm1}{Theorem}

\newcommand{\SKIP}[1]{}
\newcommand{\calV}{\mathcal{V}}
\newcommand{\calE}{\mathcal{E}}
\newcommand{\calL}{\mathcal{L}}
\newcommand{\calO}{\mathcal{O}}
\newcommand{\calG}{\mathcal{G}}
\newcommand{\hcalV}{\hat{\calV}}
\newcommand{\hcalE}{\hat{\calE}}
\newcommand{\hcalG}{\hat{\calG}}

\def\NOTE#1{{\bf [NOTE:} {\it\color{blue}{#1}}{\bf ]}.}%

\definecolor{todocolor}{rgb}{0.8,0.8,1.0}
\definecolor{fixcolor}{rgb}{1,0.8,0.8}
\definecolor{commentcolor}{rgb}{0.8,1.0,0.8}
\usepackage[color=todocolor, colorinlistoftodos]{todonotes}

\newcommand{\ignore}[1]{}

\usepackage{xspace}
\makeatletter
\DeclareRobustCommand\onedot{\futurelet\@let@token\@onedot}
\def\@onedot{\ifx\@let@token.\else.\null\fi\xspace}

\def\eg{\emph{e.g}\onedot} 
\def\ie{\emph{i.e}\onedot}

\makeatother

\begin{document}
%
\title{Memory Efficient Max Flow for Multi-label Submodular MRFs}
%
%
%
%

\author{Thalaiyasingam~Ajanthan,~\IEEEmembership{Student~Member,~IEEE,}
        Richard~Hartley,~\IEEEmembership{Fellow,~IEEE,}
        and~Mathieu~Salzmann,~\IEEEmembership{Member,~IEEE}
\IEEEcompsocitemizethanks{\IEEEcompsocthanksitem T. Ajanthan and R.
Hartley are with the College of Engineering and Computer Science,
Australian National University, Canberra and Data61, CSIRO, Canberra.\protect\\
E-mail: thalaiyasingam.ajanthan@anu.edu.au
\IEEEcompsocthanksitem M. Salzmann is with the Computer Vision Laboratory,
{\'E}cole Polytechnique F{\'e}d{\'e}rale de Lausanne.\protect
\IEEEcompsocthanksitem Data61 (formerly NICTA) is funded by the Australian
Government as represented by the Department of Broadband, Communications and the Digital Economy
and the Australian Research Council (ARC) through the ICT Centre of
Excellence program.}%
}

\IEEEtitleabstractindextext{%
\begin{abstract}
Multi-label submodular Markov Random Fields (MRFs) 
have been shown to be solvable using max-flow based on an encoding of the labels proposed by Ishikawa, in which
each variable $X_i$ is represented by $\ell$ nodes (where $\ell$ is the number of labels) arranged in a
column. However, this method in general requires $2\,\ell^2$ edges for
each pair of neighbouring variables. This makes it inapplicable to realistic problems with
many variables and labels, due to excessive memory requirement.
In this paper, we introduce a variant of the max-flow algorithm that requires much less storage.
Consequently, our algorithm makes it possible to optimally solve multi-label submodular problems involving large numbers of
variables and labels on a standard computer. 
\end{abstract}

}

\maketitle

\IEEEdisplaynontitleabstractindextext

%
\IEEEpeerreviewmaketitle

\IEEEraisesectionheading{\section{Introduction}\label{sec:introduction}}


Ishikawa~\cite{ishikawa2003exact} introduced a max-flow-based method to globally minimize the energy
 of multi-label MRFs with convex edge terms. In~\cite{schlesinger2006transforming}, this method was extended to energy functions
satisfying the \textit{multi-label submodularity} condition, analogous to the
submodularity condition for MRFs with binary labels. In the general case,
however, this method requires $2\,\ell^2$ directed edges for each pair of
neighbouring variables. For instance, for a $1000 \times 1000$, 4-connected image
with 256 labels, it would require approximately $1000\times 1000 \times 2 \times
256^2\times 2 \times 4 \approx 1000$ GB of memory to store the edges (assuming 4
bytes per edge).
Clearly, this is beyond the storage capacity of most computers.

In this paper, we introduce a variant of the max-flow algorithm that requires storing only two
$\ell$-dimensional vectors per variable pair instead of the $2\,\ell^2$ edge
capacities of the standard max-flow algorithm. In the example discussed above,
our algorithm would therefore use only 4 GB of memory for the edges. As a
result, our approach lets us optimally solve much larger problems.

More specifically, in contrast to the usual augmenting path algorithm~\cite{ford1962flows}, 
we do not store the residual edge capacities at each iteration.
Instead, our algorithm records two $\ell$-dimensional flow-related quantities for every pair 
of neighbouring variables. We show that, at any stage of the algorithm, the residual
edge capacities can be computed from these flow-related quantities and the initial edge capacities. 
This, of course, assumes that the initial capacities can be computed by some
memory-efficient routine, which is almost always the case in computer vision.


The optimality of Ishikawa's formalism made it a method of choice as a 
subroutine in many approximate energy minimization algorithms, such as multi-label
moves~\cite{torr2009improved,veksler2012multi} and
IRGC~\cite{Ajanthan_2015_CVPR}. Since our approach can simply replace the standard max-flow 
algorithm~\cite{boykov2004experimental} in Ishikawa-type graphs, 
it also allows us to minimize the energy of much larger non-submodular 
MRFs in such approximate techniques. Furthermore, due to the
similarity to standard max-flow, our algorithm can easily be extended to handle
dynamic MRFs~\cite{kohli2005efficiently} and also be accelerated using the parallel max-flow 
technique~\cite{strandmark2010parallel}.

We demonstrate the effectiveness of our algorithm on the problems of stereo
correspondence estimation and image inpainting. Our experimental evaluation
shows that our method can solve much larger problems than standard max-flow on a standard computer
and is an order of magnitude faster than state-of-the-art
message-passing
algorithms~\cite{kolmogorov2006convergent,komodakis2011mrf,SavchynskyyUAI2012}.
Our code is available at \url{https://github.com/tajanthan/memf}.

A preliminary version of this paper is appeared in~\cite{ajanthan2016memfcvpr}.
This extended version contains a polynomial time version of the MEMF algorithm,
a discussion on the equivalence with min-sum message passing and an 
experiment to evaluate the empirical time complexity.

\section{Preliminaries}
\label{sec:prelim}
Let $X_i$ be a random variable taking label $x_i \in \mathcal{L}$.
A pairwise MRF defined over a set of such random variables can be represented by an energy of the form
\begin{equation}
E(\mathbf{x}) = \sum_{i \in \mathcal{V}} \theta_i(x_i) + \sum_{(i, j) \in
\mathcal{E}} \theta_{ij}(x_i, x_j)\ ,
\label{eqn:e}
\end{equation}
where $\theta_i$ and $\theta_{ij}$ denote the unary potentials (\ie,
data costs) and pairwise potentials (\ie, interaction costs), respectively.
Here, $\mathcal{V}$ is the set of vertices, \eg, corresponding to pixels or
superpixels in an image, and $\mathcal{E}$ is the set of edges in the MRF,
\eg, encoding a 4-connected or 8-connected grid over the image pixels. 

In this work, we consider a pairwise MRF with an ordered
label set $\mathcal{L} = \{0, 1, \cdots, \ell -1 \}$, and we assume that the
pairwise terms satisfy the \textit{multi-label submodularity}
condition~\cite{schlesinger2006transforming}:
\begin{equation}
\theta_{ij}(\lambda',\mu) + \theta_{ij}(\lambda,\mu') -
\theta_{ij}(\lambda,\mu) - \theta_{ij}(\lambda',\mu') \ge 0\ ,
\end{equation}
for all $\lambda, \lambda', \mu, \mu' \in \mathcal{L}$, where $\lambda <
\lambda'$ and $\mu < \mu'$. 
Furthermore, we assume that the pairwise potentials can be computed either
by some routine or can be stored in an efficient manner.
In other words, we assume that we do not need to store each individual pairwise term. Note that, 
in computer vision, this comes at virtually no loss of generality.

\begin{figure}
\begin{center}
\includegraphics[width=0.6\linewidth, trim=4.2cm 6.3cm 10.0cm 0.8cm, clip=true,
page=70]{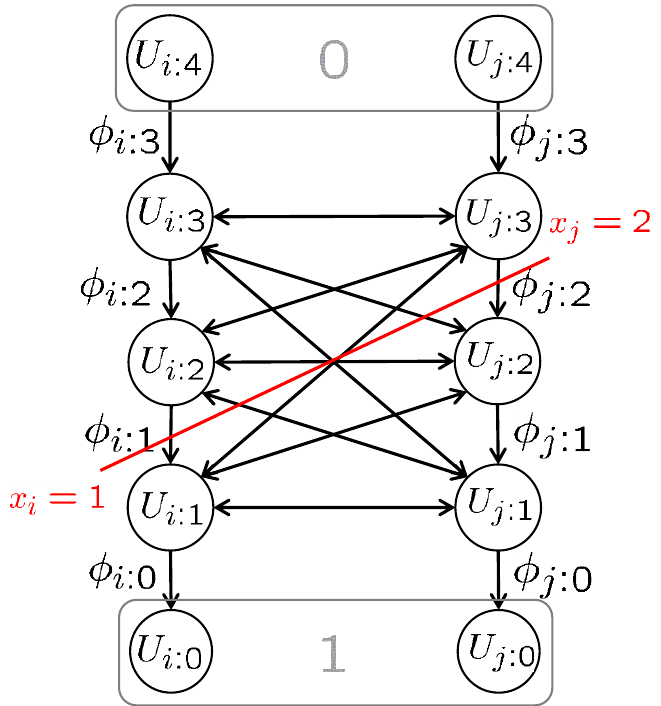}	
\end{center}
\vspace{-2ex}
\caption{\em Example of an Ishikawa graph. The graph incorporate edges with infinite capacity
from $U_{i:\lambda}$ to $U_{i:\lambda+1}$, not shown in the graph.
Here the cut corresponds to the labeling $\mathbf{x} = \{1, 2\}$ where
the label set $\calL = \{0, 1, 2, 3\}$.}
\label{fig:ish}
\vspace{-3ex}
\end{figure}

\subsection{The Ishikawa Graph}
Ishikawa \cite{ishikawa2003exact} introduced a method to represent the
multi-label energy function (\ref{eqn:e}) in a graph. The basic idea behind the
Ishikawa construction is to encode the label $X_i = x_i$ of a vertex $i\in
\mathcal{V}$ using binary-valued random variables $U_{i:\lambda}$, one for each
label $\lambda \in \mathcal{L}$.
In particular, the encoding is defined as $u_{i:\lambda} = 1$ if and only
if $x_i \ge \lambda$, and $0$ otherwise. The Ishikawa graph is then
an $st$-graph 
$\hcalG = (\hcalV\cup \{0,1\}, \hcalE)$\footnote{Some
authors denote the nodes 0 and 1 by $s$ and $t$.}, where the set of nodes and
the set of edges are defined as follows:
\begin{align}
\hcalV &= \{U_{i:\lambda} \mid i\in \calV, \lambda \in \{1, \cdots, \ell-1\}\}\
,\\\nonumber
\hcalE &= \hcalE_v \cup \hcalE_c\ ,\\\nonumber
\hcalE_v &= \{(U_{i:\lambda}, U_{i:\lambda\pm 1})\mid i\in \calV, \lambda \in
\{1, \cdots, \ell-1\}\}\ ,\\\nonumber
\hcalE_c &= \{(U_{i:\lambda}, U_{j:\mu}), (U_{j:\mu},U_{i:\lambda})\mid
(i,j)\in \calE, U_{i:\lambda}, U_{j:\mu}\in \hcalV\}\ ,
\end{align}
where $\hcalE_v$ is the set of vertical edges and $\hcalE_c$ is the set of cross
edges. Note that, the nodes $U_{i:\ell}$ and $U_{i:0}$ are identified as
node 0 and node 1 respectively. 
We denote the Ishikawa edges by $e_{ij:\lambda\mu}\in \hcalE$ (contains edges
in both directions) and their capacities by $\phi_{ij:\lambda\mu}$. We also
denote by $e_{i:\lambda}$ the downward edge $(U_{i:\lambda+1}, U_{i:\lambda})$.
An example of an Ishikawa graph is shown in Fig.~\ref{fig:ish}.

In an $st$-graph, a labeling $\mathbf{x}$ is represented by a ``cut'' in the
graph (a ``cut'' partitions the nodes in the graph into two disjoint subsets
$\hcalV_0$ and $\hcalV_1$, with $0 \in \hcalV_0$ and $1 \in
\hcalV_1$). Then, the value of the energy function $E(\mathbf{x})$ is equal to the sum
of the capacities on the edges from $\hcalV_0$ to $\hcalV_1$. 
In an Ishikawa graph, if the downward edge $e_{i:\lambda}$ is in the ``cut'',
then vertex $i$ takes label $\lambda$.
In MRF energy minimization, each vertex $i$ takes exactly one label $x_i$, which means that exactly 
one edge $e_{i:\lambda}$ must be in the min-cut
of the Ishikawa graph. This is ensured by having infinite capacity for each
upward edge $e_{ii:\lambda\lambda+1}$, \ie,
$\phi_{ii:\lambda\lambda+1} = \infty$ for all $i\in \calV$ and $\lambda\in
\calL$.
Note that, by construction of the Ishikawa graph, the capacities $\phi$ and
the energy parameters $\theta$ are related according to the following formula:
\begin{align}
\label{eqn:thph0}
\theta_{i}(\lambda) &= \phi_{ii:\lambda+1\lambda} = \phi_{i:\lambda}\
,\\\nonumber
\theta_{ij}(\lambda,\mu) &= \sum_{\substack{\lambda' > \lambda\\ \mu' \le \mu}}
\phi_{ij:\lambda'\mu'} + \sum_{\substack{\lambda' \le \lambda\\ \mu' > \mu}}
\phi_{ji:\mu'\lambda'}\ .
\end{align}

Finding the minimum energy labeling is a min-cut problem, which can be
solved optimally using the max-flow algorithm~\cite{ford1962flows} when the
edge capacities are non-negative. As shown in~\cite{schlesinger2006transforming}, a \textit{multi-label
submodular} energy function can be represented by an Ishikawa graph with
non-negative edge capacities $\phi$ and can therefore be minimized optimally
by max-flow.

\subsection{Max Flow}
The most popular max-flow algorithm in computer
vision~\cite{boykov2004experimental} is an augmenting path algorithm that finds
a path from node 0 to node 1 through positive edges (called an \textit{augmenting path}) and then pushes the 
maximum flow without exceeding the edge capacities
(called \textit{augmentation}). The augmentation operation changes the edge
capacities in the graph, and therefore, the residual graph needs to be stored.
That is, when applied to the Ishikawa graph, the max-flow algorithm stores
$2\,\ell^2$ values per pair of neighbouring variables. For large number of
labels and of variables, the memory requirement is high and, in many practical problems, 
exceeds the capacity of most computers.

\subsection{Our Idea}\label{sec:idea}
Let us assume that the max-flow algorithm is applied to
the Ishikawa graph. As the algorithm proceeds, the capacities on the edges in the graph
change in response to the flow. Here, instead of storing the
residual graph, we propose recording the flow that has been applied to the
graph.

However, since storing the flow would also require $2\,\ell^2$ values per
variable pair, we propose recording two $\ell$-dimensional quantities related to
the flow between pair of variables.
More precisely, for each directed edge\footnote{$\mathcal{E}^+$ denotes 
the set of directed edges between the vertices in the MRF,
 \ie, if $(i,j)\in\mathcal{E}$ then, $(i,j)\in \mathcal{E}^+$ and $(j,i)\in \mathcal{E}^+$.}
  $(i,j)\in\mathcal{E}^+$, we record the sum of outgoing
flows from each node $U_{i:\lambda}$ to the nodes $U_{j:\mu}$ for all $\mu \in
\{1,\cdots,\ell-1\}$. We call this quantity an \textit{exit-flow}, denoted by
$\Sigma_{ij:\lambda}$ (defined below in Eq.~\ref{eqn:sig}). We show that these
exit-flows allow us to reconstruct a \textit{permissible} flow (defined below
in Def.~\ref{dfn:pflow}), which in turn lets us compute the residual edge
capacities from the initial ones. Importantly, while flow reconstruction is not unique, 
we show that all such reconstructions
are equivalent up to a \textit{null} flow (Def.~\ref{dfn:cflow}), which does not affect the
energy function. Note that this idea can be applied to any augmenting path
algorithm, as long as the residual graph can be rapidly constructed.

For increased efficiency, we then show how finding an augmenting path can be
achieved   
in a simplified Ishikawa graph (called \textit{block-graph}) that amalgamates
the nodes in each column into blocks.
We then perform augmentation, which translates to updating our
exit-flows, in this block-graph.
As a side effect, since an augmenting path in our block-graph 
corresponds to a collection of augmenting paths in the Ishikawa graph, our algorithm converges 
in fewer iterations than the standard max-flow implementation
of~\cite{boykov2004experimental}.



\section{Memory efficient flow encoding}\label{sec:encode}
Before we introduce our memory efficient max flow algorithm, let us
describe how the cumulative flow can be stored in a memory efficient manner.
This technique can be used in any augmenting path flow algorithm, by
reconstructing the residual edge capacities whenever needed.

Let us assume that the max-flow algorithm is applied to
the Ishikawa graph. At some point in the algorithm, flow has passed along many of the edges of the
graph.


\begin{dfn}\label{dfn:flow}
A flow is a mapping $\psi: \hcalE \to {\rm
I\!R}$, denoted by $\psi_{ij:\lambda\mu}$
for the edges $e_{ij:\lambda\mu}$, 
that satisfies the anti-symmetry condition $\psi_{ij:\lambda\mu} =
-\psi_{ji:\mu\lambda}$ for all $e_{ij:\lambda\mu}\in \hcalE$. 

A flow is called
\textit{conservative}\footnote{A conservative flow is often
referred to as a flow in the literature.} if the total flow into a node is zero
for all nodes, except for the source and the terminal, \ie,
\begin{equation}\label{eqn:fcon}
\sum_{j,\mu\mid e_{ji:\mu\lambda}\in \hcalE} \psi_{ji:\mu\lambda} = 0 \quad
\forall\, U_{i:\lambda}\in \hcalV\ .
\end{equation}
%

Given $\psi$, the residual capacities of the Ishikawa graph
are updated as
 $\phi = \phi^0 - \psi$, where $\phi^0$ represents the initial
edge capacities. 
Furthermore, we call the flow restricted to each column 
\textit{column-flows}, which we denote by
$\psi_{i:\lambda}\,;\,i\in \calV, \lambda\in \calL$.
\end{dfn}

At first sight, it might seem that, to apply the max-flow algorithm, it is
necessary to keep track of all the values $\psi_{ij:\lambda\mu}$, which would
require the same order of storage as recording all the edge capacities. 
Below, however, we show that it is necessary
to store only $\mathcal{O}(\ell)$ values for each $(i,j)\in\mathcal{E}$, 
instead of $\mathcal{O}(\ell^2)$. 

To this end, the flow values that we store in our algorithm, namely
\textit{source-flows} and \textit{exit-flows} are defined below.
\begin{dfn}
\begin{enumerate}
  \item For each $i \in \calV$, the flow out from the source node
 $\psi_{i:\ell-1}$ is called a \textit{source-flow}.
  \item For each $(i,j)\in\mathcal{E}^+$ and $\lambda \in \{1, \cdots, \ell-1\}$, we
define an \textit{exit-flow} as
\begin{equation}
\label{eqn:sig}
\Sigma_{ij:\lambda} = \sum_{\mu} \psi_{ij:\lambda\mu}\ .
\end{equation}
\end{enumerate}
\end{dfn}

%
%
We will show that these source-flows and exit-flows permit the flow
$\psi$ to be reconstructed up to equivalence.

Now, let us define some additional properties of flow, which will be useful in
our exposition.
\begin{dfn}\label{dfn:pflow}
A flow $\psi$ is called \textit{permissible} if $\phi^0_{ij:\lambda\mu} - 
\psi_{ij:\lambda\mu} \ge 0$ for all $e_{ij:\lambda\mu}\in \hcalE$.
\end{dfn}

\begin{dfn}\label{dfn:cflow}
A flow $\psi$ is called \textit{null} if the total flow into a node is zero 
for all nodes including the source and the terminal, \ie, satisfies
Eq.~\ref{eqn:fcon} for all $U_{i:\lambda}\in \hcalV \cup\{0,1\}$.

Note that a null flow does not change the energy function
represented by the $st$-graph and it is identical to passing flow around loops.
Also, if $\psi$ is a null flow then so is $-\psi$.
\end{dfn}

Furthermore, note that the energy function encoded by an $st$-graph is a
quadratic pseudo-boolean function~\cite{boros2002pseudo}, and a
\textit{reparametrization} of such a function is identical to a null
flow in the corresponding $st$-graph. 

\begin{lem}\label{lem:eqflow} 
Two sets of capacities $\phi$ and $\phi'$ represent the same energy
function exactly (not up to a constant), written as $E_\phi \equiv E_{\phi'}$, 
if and only if $\phi' - \phi$ is a null
flow.
\end{lem}
\begin{proof}
This lemma is a restatement of the reparametrization lemma
of~\cite{kolmogorov2006convergent,werner2007linear} in the context of
$st$-graphs.
\end{proof}

\begin{figure}
\begin{center}
\begin{subfigure}{0.33\linewidth}
\begin{center}
\includegraphics[width=0.9\linewidth, trim=2.8cm 7.7cm 14.0cm 3.9cm,
clip=true, page=72]{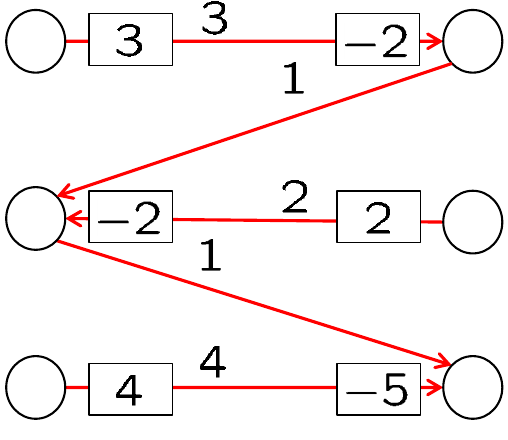}	
\vspace{-3ex}
\end{center}
\caption{$\psi$}
\end{subfigure}%
\begin{subfigure}{0.33\linewidth}
\begin{center}
\includegraphics[width=0.9\linewidth, trim=2.8cm 7.7cm 14.0cm 3.9cm,
clip=true, page=71]{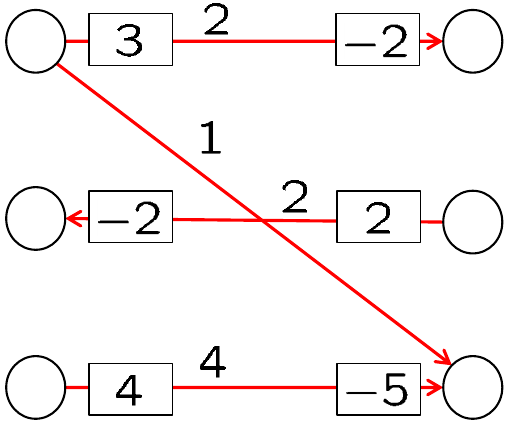}	
\vspace{-3ex}
\end{center}
\caption{$\psi'$}
\end{subfigure}%
\begin{subfigure}{0.33\linewidth}
\begin{center}
\includegraphics[width=0.9\linewidth, trim=2.8cm 7.7cm 14.0cm 3.9cm,
clip=true, page=73]{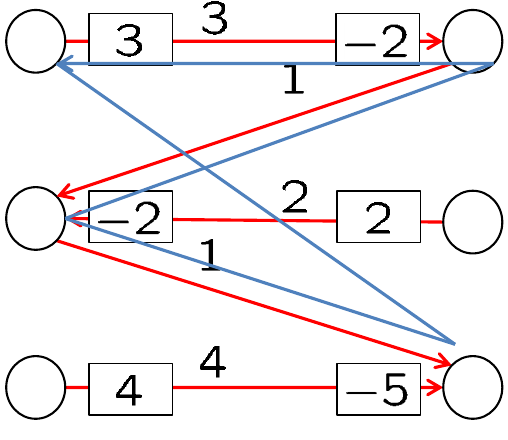}	
\vspace{-3ex}
\end{center}
\caption{$\psi \equiv \psi'$}
\end{subfigure}
\end{center}
\vspace{-3ex}
\caption{\em An example of two equivalent flow representations with the same
exit-flows.
Note that each red arrow represents the value $\psi_{ij:\lambda\mu}$ and the
opposite arrows $\psi_{ji:\mu\lambda}$ are
not shown. Furthermore,  
the exit-flows $\Sigma$ are shown next to
the nodes and the initial edges $\phi^0$ are not shown. 
In \textbf{(c)}, the flow $\psi'$ is obtained from $\psi$ 
by passing flow around a loop.}
\label{fig:rec}
\vspace{-3ex}
\end{figure}

Let $\phi$ and $\phi'$ be two sets of residual capacities obtained from an
initial set of capacities $\phi^0$ by passing two flows $\psi$ and $\psi'$, 
\ie, $\phi = \phi^0 - \psi$
and $\phi' = \phi^0 - \psi'$. If $\phi$ and $\phi'$ are equivalent, then, by
Lemma~\ref{lem:eqflow}, $(\phi^0 - \psi) - (\phi^0 - \psi') = \psi' - \psi$ is a
null flow.
Hence $\psi'$ can be obtained from $\psi$ by passing flow around loops in the graph.
See Fig.~\ref{fig:rec}.

We can now state our main theorem.

\begin{thm}\label{thm:fe}
Let $\phi^0$ be the initial capacities of an Ishikawa graph, and let
$\Sigma$ 
be a set of exit-flows.  Suppose that $\psi$ and $\psi'$
are two conservative flows compatible with $\Sigma$, meaning that
(\ref{eqn:sig}) holds for both $\psi$ and $\psi'$, and that $\psi$ and $\psi'$
have identical source-flows.
Then $E_{\phi^0 - \psi} \equiv E_{\phi^0 - \psi'}$.
\end{thm}
%

The idea is then as follows.  If a permissible conservative flow $\psi$ is
obtained during an augmenting path flow algorithm, but only the exit-flows 
$\Sigma_{ij:\lambda}$ are retained for each $(i, j)\in \calE^+$ and label $\lambda$,
then one wishes, when required, to reconstruct the flow $\psi$ on a given
edge $(i, j)\in \calE$.  
Although the reconstructed flow $\psi'$ may not
be identical with the flow $\psi$, the two will result in equivalent energy
functions (not just equal up to a constant, but exactly equal for all
assignments).
In the augmenting path algorithm, the current flow values are only needed
temporarily, one edge at a time, to find a new augmenting path, and
hence do not need to be stored, as long as they can be rapidly computed. 

Now we prove Theorem~\ref{thm:fe}.
\begin{proof}
 \vspace{-0.2cm}
First we will prove that $\psi$ and $\psi'$ have identical column-flows. For a
conservative flow
\begin{equation}
\label{eqn:colf}
\psi_{i:\lambda} - \left(\psi_{i:\lambda-1} + \sum_{(i,j)\in \calE^+}
\Sigma_{ij:\lambda}\right) = 0\ ,
\end{equation}
for all $i\in \calV$ and $\lambda\in\{1,\ldots,\ell-1\}$. Since $\psi$ and
$\psi'$ are compatible with $\Sigma$ and have identical source-flows,
$\psi_{i:\lambda} = \psi'_{i:\lambda}$ for all $i\in \calV$ and $\lambda =
\calL$. Hence they have identical column-flows. 

Now we will prove the equivalence.
Given a flow $\psi$, let us denote its restriction to the edges
$e_{ij:\lambda\mu}$ for all $\lambda,\mu\in\{1,\ldots,\ell-1\}$ for some
$(i,j)\in \calE$ by $\psi_{ij}$, \ie restriction to cross edges only.
Since both $\psi_{ij}$ and $\psi'_{ij}$ satisfy Eq.~\ref{eqn:sig}, $\psi'_{ij} - \psi_{ij}$ is a null flow. 
Furthermore, since both $\psi$ and $\psi'$ have identical column-flows, 
$\psi' - \psi = (\phi^0 - \psi) - (\phi^0 - \psi')$ is a null flow and, by
Lemma~\ref{lem:eqflow}, $E_{\phi^0 - \psi} \equiv E_{\phi^0 - \psi'}$.
\end{proof}

\SKIP{
\begin{algorithm}[t]
\caption{Flow reconstruction}
\label{alg:fe}
\begin{algorithmic}
\Require Given a directed edge $(i,j)\in\mathcal{E}^+$
\For{$\lambda \gets \ell-1$ \textbf{to} $1$}
\If{$\Sigma_{ij:\lambda} \ge 0$}
\For{$\mu \gets \ell-1$ \textbf{to} $1$}
\If{$\Sigma_{ji:\mu} \le 0$}
\State $\psi'_{ij:\lambda\mu} \gets \min(|\Sigma_{ij:\lambda}|,
	\phi^0_{ij:\lambda\mu}, |\Sigma_{ji:\mu}|)$
\State $\psi'_{ji:\mu\lambda} \gets -\psi'_{ij:\lambda\mu}$
\State $\Sigma_{ji:\mu} \gets \Sigma_{ji:\mu} - \psi'_{ji:\mu\lambda}$
\State $\Sigma_{ij:\lambda} \gets \Sigma_{ij:\lambda} - \psi'_{ij:\lambda\mu}$
\If{$\Sigma_{ij:\lambda} = 0$}
\State \textbf{break}
\EndIf
\EndIf
\EndFor
\EndIf
\EndFor
\end{algorithmic}
\end{algorithm}
}

\begin{figure}[t]
\vspace{-0.3cm}
\begin{center}
\includegraphics[width=0.95\linewidth, trim=0.8cm 9.4cm 2.5cm 3.3cm, clip=true,
page=82]{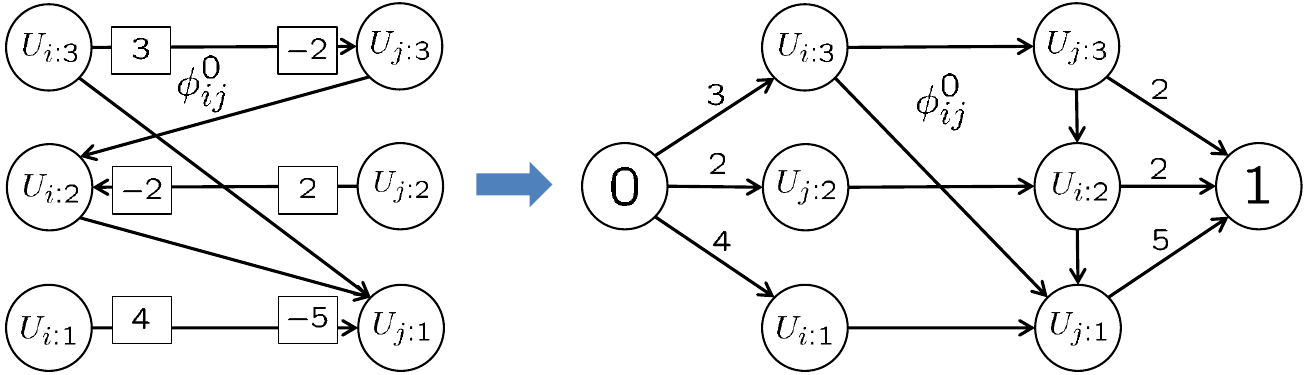}	
\vspace{-3ex}
\end{center}
\caption{\em Given $\phi^0$ and $\Sigma$ (left), flow reconstruction is
formulated as a max-flow problem (right). Here the nodes with positive
exit-flows are connected to the source (0) and those with negative exit-flows
are connected to the terminal (1).}
\label{fig:frec}
\vspace{-3ex}
\end{figure}

\vspace{-0.4cm}
\subsection{Flow Reconstruction}\label{sec:frec}
Note that, from Eq.~\ref{eqn:colf} it is clear that given the source-flows
$\psi_{i:\ell-1}\,;\,i\in \calV$, the column-flows $\psi_{i:\lambda}\,;\,i\in
\calV, \lambda\in \calL$ can be computed in a top-down fashion. Now we
turn to the problem of finding the flows along the cross-edges
$e_{ij:\lambda\mu}$. 

Given the set of exit-flows $\Sigma$, the objective
 is to find a permissible flow $\psi'$
satisfying Eq.~\ref{eqn:sig}. Note that there exists a permissible
conservative flow $\psi$ compatible with $\Sigma$ and hence we find $\psi'$ such
that $\psi' - \psi$ is a null flow. We do this by considering one edge $(i,j)\in \calE$ at a time
and reconstruct the flow by formulating a small max-flow problem.

Considering all the
nodes $U_{i:\lambda}$ and $U_{j:\mu}$ for a given pair $(i,j)$, we join them with
edges with initial capacities $\phi^0_{ij:\lambda\mu}$. Nodes with
positive exit-flow $\Sigma_{ij:\lambda}$ are joined to the source with edges
of capacities $\left|\Sigma_{ij:\lambda}\right|$. Similarly, those with negative
exit-flow are joined to the terminal. See Fig.~\ref{fig:frec}. 

Note that, in this network, the edges
from the source can be thought of as ``supply'' and the edges to
the terminal can be thought of as ``demand''.
Since the total supply equals the total demand in this network and there exists
a permissible flow $\psi_{ij}$ compatible with $\Sigma$ (\ie, satisfying the
supply-demand equality), the maximum flow solution of this network $\psi'_{ij}$
is compatible with $\Sigma$, \ie, satisfies Eq.~\ref{eqn:sig}. In fact we are
interested in non-negative residual capacities $\phi'_{ij} = \phi^0_{ij} -
\psi'_{ij}$ which are readily available in this network.
This problem can be solved using a greedy augmenting path
algorithm. While this graph has $\calO(\ell)$ nodes and $\calO(\ell^2)$ edges,
this remains perfectly tractable, since we only consider one edge $(i,j)$ at a
time. Therefore, ultimately, flow reconstruction can be done efficiently.

At this point, given the initial capacities $\phi^0$, the source-flows
$\psi_{i:\ell-1}\,;\, i\in \calV$ and the set of exit-flows $\Sigma$,
we have shown how to reconstruct the
non-negative residual edge capacities $\phi'$. 
This requires $\calO(|\calV| + |\calE|\,\ell)$ values to be stored.


\section{Polynomial Time Memory Efficient Max Flow}\label{sec:pmemf}
We now introduce our polynomial time memory efficient max flow algorithm, which
minimizes multi-label submodular MRF energies with pairwise interactions.
Our algorithm follows a similar procedure as the standard Edmonds-Karp
algorithm~\cite{thomas2001introduction}, in that it iteratively finds the
shortest augmenting path and then pushes the maximum flow through it
without exceeding the edge capacities.
However, instead of storing the residual graph,
we store exit-flows as proposed in Section~\ref{sec:encode}, which, at any stage
of the algorithm, would allow us to compute the residual graph. Below, we discuss
how one can find an augmenting path and update the exit-flows,
\ie, perform augmentation, without storing the full Ishikawa graph.

\vspace{-0.4cm}
\subsection{Finding an Augmenting Path}\label{sec:pfind}
Our algorithm finds an augmenting path in a subgraph of the Ishikawa graph,
called \textit{lower-graph}.
In particular, the lower-graph contains only a subset of Ishikawa edges which
satisfy the \textit{lowest-cross-edge} property.
\begin{dfn}
Consider a directed edge $(i,j)\in \calE^+$. For each node
$U_{i:\lambda}$, the \textit{lowest-cross-edge} is defined as, the edge
$e_{ij:\lambda\mu}$ where $\mu$ is the smallest value such that
$\phi_{ij:\lambda\mu} > 0$.
\end{dfn}
More specifically, in addition to the vertical edges $\hcalE_v$, the lower-graph
contains the lowest-cross-edges. Therefore, we only store $\mathcal{O}(\ell)$
edges per variable pair $(i,j)$. Now, the relationship between augmenting paths
in the original Ishikawa graph and the lower-graph can be characterized by the
following theorem.
\begin{thm}\label{thm:low}
\vspace{-0.1cm}
Given the Ishikawa graph, there is an augmenting path in the lower-graph if and
only if there exists an augmenting path in the Ishikawa graph.
\end{thm}
\begin{proof}
\vspace{-0.1cm}
Since the lower-graph is a subgraph of the Ishikawa graph, if there is an
augmenting path in the lower-graph, then there exists an augmenting path in the
Ishikawa graph. 

We will now prove the converse. Consider a directed edge $(i,j)\in \calE^+$. Let
$e_{ij:\lambda\mu}$ and $e_{ij:\lambda\mu'}$ be two positive capacity edges
from $U_{i:\lambda}$ and $e_{ij:\lambda\mu'}$ be the lowest-cross-edge. Then,
due to the upward infinite capacity edges from $U_{j:\mu'}\leadsto U_{j:\mu}$, there is
a positive capacity path from $U_{i:\lambda}\leadsto U_{j:\mu}$ through the
lowest-cross-edge $e_{ij:\lambda\mu'}$. This proves the theorem.
\vspace{-0.1cm}
\end{proof}

This enables us to find all the augmenting paths in the Ishikawa graph by
searching in a smaller graph that has $\mathcal{O}(\ell)$ edges per variable
pair $(i,j)$.


Note that, as mentioned earlier, we find the \textit{shortest} augmenting path
in this lower-graph. However, by contrast to the Edmonds-Karp
algorithm~\cite{thomas2001introduction}, the path distance is computed
considering \textit{zero} distance for the infinite capacity edges and unit distance for 
other edges, instead of unit distance for all the edges. 
The intuition for this modification is that, the infinite capacity edges
will never become saturated (or eliminated from the graph) for the entire
course of the algorithm. Note that, with this definition of path distance, the
augmenting paths in both lower-graph and Ishikawa graph have same length. This
will enable us to prove the polynomial time bound of our algorithm in a 
similar manner as the standard Edmonds-Karp algorithm.
Note that, even in this case, the shortest augmenting path can be
found using a Breadth First Search (BFS) scheme. 

\subsection{Augmentation}\label{sec:paug}
Now, given an augmenting path $p$, we want to push the maximum permissible flow
through it. The edges in the augmenting path $p$ are updated in the similar
manner as in the usual max-flow algorithm. In addition to that, for each
cross edge $e_{ij:\lambda\mu}\in \hcalE_c$ that is in the augmenting path, the
exit-flows are updated as follows:
\vspace{-0.2cm}
\begin{align}
\Sigma_{ij:\lambda} &= \Sigma_{ij:\lambda} + \alpha\ ,\\\nonumber
\Sigma_{ji:\mu} &= \Sigma_{ji:\mu} - \alpha\ ,
\end{align}
where $\alpha$ is the maximum possible flow along the path $p$.

After the flow augmentation, the lower-graph needs to be updated to maintain
the lowest-cross-edge property.
Note that the lowest-cross-edge property may be violated due to the following reasons:
\begin{enumerate}
  \item A new lowest-cross-edge $e_{ij:\lambda\mu}$ is created due
  to a flow along the edge $e_{ji:\mu\lambda}$.
  \item A new lowest-cross-edge $e_{ij:\lambda\mu}$ is created due
  to a saturating flow along $e_{ij:\lambda\mu'}$ for some $\mu' <
  \mu$, \ie, the edge $e_{ij:\lambda\mu'}$ disappears from the Ishikawa graph.
\end{enumerate}

Note that, during an augmentation, if a new lowest-cross-edge is created due to
a flow in the opposite direction (case-1 above), then the new lowest-cross-edge is known and
the lower-graph can be updated directly, \ie, the new lowest-cross-edge can be
stored.

On the other hand, if a cross edge becomes saturated (case-2), then we
need to run the flow reconstruction algorithm to find the new lowest-cross-edge 
and update the lower-graph. This can be done in a memory efficient manner,
since it only involves one edge $(i,j)\in \calE$ at a time. 

\begin{algorithm}[t]
\caption{Memory Efficient Max Flow (MEMF) - Polynomial Time Version}
\label{alg:pmemf}
\begin{algorithmic}

\Require $\phi^0$ \Comment{Initial Ishikawa capacities}

\State $\Sigma \gets 0$ \Comment{Initialize exit-flows}
\State $\bar{\phi} \gets $ lower-graph($\phi^0$) \Comment{Store the 
lowest-cross-edges}

\Repeat 

\State $p \gets $ shortest\_augmenting\_path($\bar{\phi}$) 
\Comment{Sec.~\ref{sec:pfind}}

\State $(\bar{\phi}, \Sigma) \gets $ augment($p, \bar{\phi}$) 
\Comment{Sec.~\ref{sec:paug}}

\ForAll{edge $e_{ij:\lambda\mu}$ becomes saturated}
\State $\phi_{ij} \gets $ compute\_edges($\phi^0, \Sigma, i, j$)
\Comment{Sec.~\ref{sec:frec}}

\State $\bar{\phi}_{ij} \gets $ lower-graph($\phi_{ij}, i, j$)
\Comment{Sec.~\ref{sec:pfind}}

\EndFor

\Until no augmenting paths possible 

\\\Return get\_labelling($\bar{\phi}$)
\Comment{Find the cut using BFS}

\end{algorithmic}
\end{algorithm}

\subsection{Summary}\label{sec:spmemf}
Our polynomial time memory efficient max flow is summarized in
Algorithm~\ref{alg:pmemf}. 
Let us briefly explain the subroutines below.

\paragraph*{\textbf{lower-graph}}
Given the initial Ishikawa edge capacities $\phi^0$, this subroutine constructs
the lower-graph (with edge capacities $\bar{\phi}$) by retaining the
lowest-cross-edges from each node $U_{i:\lambda}\in \hcalV$, for each directed edge
$(i,j)\in \calE^+$ (see Sec.~\ref{sec:pfind}). If the input to
this subroutine is the Ishikawa capacities $\phi_{ij}$ corresponding to the edge
$(i,j)\in \calE$, then it retains the lowest-cross-edges $\bar{\phi}_{ij}$.

\paragraph*{\textbf{shortest\_augmenting\_path}} 
Given the lower-graph parameters $\bar{\phi}$, this subroutine finds the
shortest augmenting $p$ using BFS, as discussed in
Section~\ref{sec:pfind}.

\paragraph*{\textbf{augment}}
Given the path $p$, this subroutine finds the maximum
possible flow through the path and updates the lower-graph and the
set of exit-flows, as discussed in Section~\ref{sec:paug}. In addition, if a new
lowest-cross-edge is created due to a flow in the opposite direction (case-1 in
Sec.~\ref{sec:paug}), then it also updates the lower-graph capacities
$\bar{\phi}$.

\paragraph*{\textbf{compute\_edges}}
Given the initial Ishikawa edge capacities $\phi^0$ and the set of exit-flows
$\Sigma$, this subroutine computes the non-negative residual Ishikawa capacities
$\phi_{ij}$ corresponding to the given edge $(i,j)$. This is accomplished by
solving a small max-flow problem (see Sec.~\ref{sec:frec}).


\paragraph*{\textbf{get\_labelling}}
This subroutines finds the partition of the lower-graph by running BFS.

As discussed above, the exit-flows $\Sigma$ require
$\mathcal{O}(\ell)$ storage for each edge $(i,j)\in\mathcal{E}$. In addition,
the lower-graph can have at most
$\mathcal{O}(|\mathcal{V}|\,\ell)$ nodes and
$\mathcal{O}(|\mathcal{E}|\,\ell)$ edges. Furthermore, recall that we assume
that the initial Ishikawa edge capacities $\phi^0$ can be stored efficiently.
Therefore, ultimately, the space complexity of our algorithm is
$\mathcal{O}((|\mathcal{V}| + |\mathcal{E}|)\,\ell) =
\mathcal{O}(|\mathcal{E}|\,\ell)$. Let us now prove the polynomial time bound of
our algorithm.

\subsection{Time Complexity Analysis}
We follow the time complexity analysis of the standard Edmonds-Karp
algorithm~\cite{thomas2001introduction} to derive a polynomial time bound on
our algorithm. In particular, first the analysis proves that the shortest path
distance from source (node 0) to any node is monotonically increasing with
each flow augmentation. Then, it derives a bound on the number of
augmentations. In fact, the number of augmentations of our MEMF algorithm also
has the same bound as the Edmonds-Karp algorithm.

\begin{thm}
If the MEMF algorithm is run on the Ishikawa graph
$\hcalG=(\hcalV\cup\{0,1\}, \hcalE)$ with source $0$ and terminal $1$,
then the total number of augmentations performed by the algorithm is
$\mathcal{O}(|\hcalV||\hcalE|)$.
\end{thm}
\begin{proof}
The proof follows the steps of standard proof of the Edmonds-Karp algorithm. See
Appendix~\ref{app:poly} for details.
\end{proof}

Let us analyze the time complexity of each subroutine below.
Note that, both the subroutines \textit{shortest\_augmenting\_path} and
\textit{augment} runs in $\mathcal{O}(|\calE|\,\ell)$ time, since, in the worst
case, both subroutines need to check each edge in the lower-graph. However,
\textit{compute\_edges} required to run the flow reconstruction 
algorithm which takes $\mathcal{O}(\ell^3)$ time for each variable pair $(i,j)$ (assuming a small
max-flow problem with $2\,\ell$ nodes and $\ell^2$ edges, solved using the most
efficient algorithm~\cite{orlin2013max}, see Sec.~\ref{sec:frec}). Also
\textit{lower-graph} requires $\mathcal{O}(\ell^2)$ time for each variable
pair $(i,j)$, since it needs to check each of the Ishikawa edges. Hence, the
worst case running time of each iteration (\ie, augmentation step) is
$\mathcal{O}(|\calE|\,\ell + K\,(\ell^3 + \ell^2)) =
\mathcal{O}(|\calE|\,\ell + |\calE|\,\ell^3) = \mathcal{O}(|\calE|\,\ell^3)$,
where $K$ is the maximum number of flow-reconstructions (\ie, saturated cross
edges) at an augmentation step. Since the number of augmentations is bounded by
$\mathcal{O}(|\hcalV||\hcalE|)$, 
the worst case running time of the entire execution of the MEMF algorithm is
$\mathcal{O}(|\calV|\,\ell\,|\calE|\,\ell^2\,|\calE|\,\ell^3) =
\mathcal{O}(|\calV||\calE|^2\,\ell^6)$. This is $\mathcal{O}(\ell)$ slower
than the standard Edmonds-Karp algorithm on the Ishikawa graph. Note that,
however, MEMF requires $\mathcal{O}(\ell)$ less memory.

\section{Efficient algorithm}\label{sec:memf}
In the previous section, we have provided a general purpose polynomial time
max-flow algorithm that is also memory efficient. However, for computer vision
applications, the BK algorithm~\cite{boykov2004experimental} is shown to be
significantly faster than the standard max-flow implementations, even though it
lacks the polynomial time guarantee. The basic idea is to maintain a search
tree
throughout the algorithm instead of building the search tree from scratch at
each iteration. 

Motivated by this, we also propose doing search-tree-recycling
similar to the BK algorithm. 
Since we lose the polynomial time guarantee,
for increased efficiency, we further
simplify the Ishikawa graph. In particular, we find an augmenting path in a 
\textit{block-graph}, that amalgamates the nodes in each column into blocks.
Since an augmenting path in our block-graph corresponds to a collection of
augmenting paths in the Ishikawa graph, our algorithm converges in fewer
iterations than the BK algorithm.

\subsection{Efficiently Finding an Augmenting Path}\label{sec:find}
As mentioned above, we find an augmenting path in a block-graph\footnote{We
called this a \textit{simplified graph} in~\cite{ajanthan2016memfcvpr}.}, whose
construction is detailed below.

\begin{figure}
\begin{center}
\includegraphics[width=0.95\linewidth, trim=0.2cm 4.0cm 4.1cm 0.5cm, clip=true,
page=11]{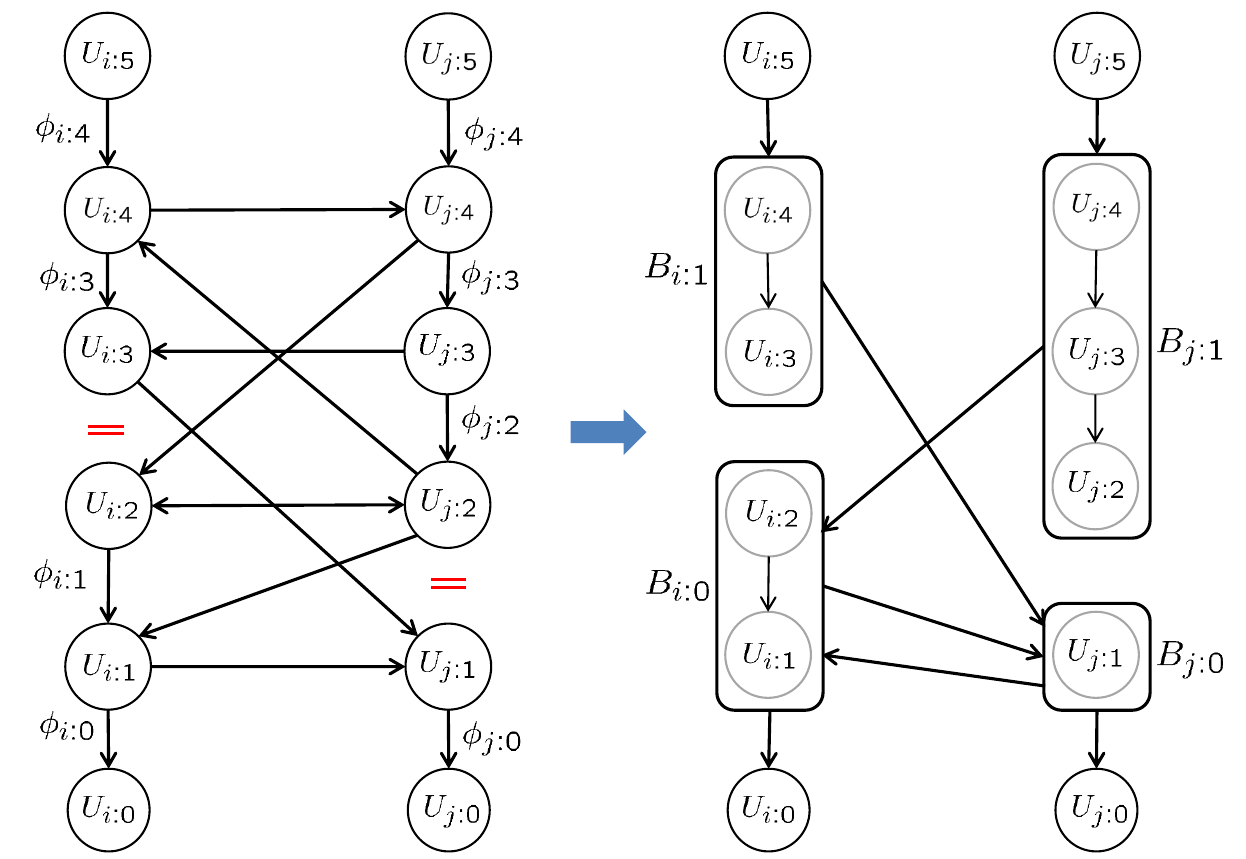}	
\end{center}
\vspace{-3ex}
\caption{\em To find an augmenting path in a memory efficient manner, 
we propose a simplified representation of the Ishikawa graph in terms of blocks
corresponding to consecutive non-zero edges in each column $i$.}
\label{fig:red}
\vspace{-3ex}
\end{figure}

Given the parameters $\phi$, we rely on the fact that there exists a label $\lambda$ such that
$\phi_{i:\lambda} = 0$ for each $i\in\mathcal{V}$. In fact, it is easy to see
that in each column $i$, if all $\phi_{i:\lambda}$ are positive, then there
exists a \textit{trivial} augmenting path from $U_{i:\ell}$ to $U_{i:0}$, and the
minimum along the column can be subtracted from each $\phi_{i:\lambda}$. Now, at each column
$i$, we partition the nodes $U_{i:\lambda}$ for all $\lambda \in \{1, \cdots, \ell-1\}$
into a set of \textit{blocks}, such that each node in a block is connected with
positive edges $\phi_{i:\lambda}$. Let us denote these blocks by $B_{i:\gamma}$, where
$\gamma$ is indexed from bottom to top starting from 0.
Note that there is no edge between $B_{i:\gamma}$ and $B_{i:\gamma\pm1}$. 
As depicted by Fig.~\ref{fig:red}, our
block-graph then contains only the blocks and the edges between the blocks. 

The edges in the block-graph are obtained as follows.
Let us consider a directed edge $(i,j) \in\mathcal{E}^+$. We add an edge 
$B_{i:\gamma} \to B_{j:\delta}$,
where $\delta$ is the smallest value such that $\phi_{ij:\lambda\mu} >0$
 for some $U_{i:\lambda} \in B_{i:\gamma}$ and $U_{j:\mu} \in B_{j:\delta}$. 
While doing this, we also enforce that there is no edge $B_{i:\gamma'} \to
B_{j:\delta'}$ such that $\gamma' > \gamma$ and $\delta' < \delta$. 
The reasoning behind this is that, because of the upward infinite-capacity
edges between the nodes
$U_{i:\lambda}$ and $U_{i:\lambda+1}$, we have the following:
\begin{enumerate}
  \item If a node $U_{j:\mu}$ can be reached from $U_{i:\lambda}$ through
  positive edges, then the nodes $U_{j:\mu'}$, for all $\mu' \ge \mu$, can also be
  reached.
  \item If a node $U_{j:\mu}$ can be reached from $U_{i:\lambda}$ through
  positive edges, then it can also be reached from the nodes $U_{i:\lambda'}$, for all
  $\lambda' \le \lambda$.
\end{enumerate}
Hence, an edge $B_{i:\gamma} \to B_{j:\delta}$
indicates the fact that there is some positive flow possible from any node
$U_{i:\lambda} \in B_{i:\gamma'}$, for all $\gamma' \le \gamma$, to any node
$U_{j:\mu} \in B_{j:\delta'}$, for all $\delta' \ge \delta$. In other words, the 
set of edges obtained by this procedure is sufficient. 

Now, the relationship between augmenting paths in the original Ishikawa graph
and in our block-graph can be characterized by the following theorem.
\begin{thm}\label{thm:sim}
Given the set of Ishikawa graph parameters $\phi$, there is an augmenting path
in the block-graph if and only if there exists an augmenting
path in the Ishikawa graph.
\end{thm}
\begin{proof}
The basic idea of this proof is the same as the proof of Theorem~\ref{thm:low}.
See Appendix~\ref{app:sim}.
\end{proof}

Note that the block-graph can only be used to find an augmenting path; 
the quantity of the maximum permissible flow cannot be determined in this
graph. Therefore, the capacity of an edge $B_{i:\gamma} \to B_{j:\delta}$ is not
important, but it is important to have these edges. Note also that the
block-graph is constructed incrementally for each edge $(i,j)\in\mathcal{E}$. 
Hence, it only requires us to store the Ishikawa graph parameters $\phi_{ij}$ 
corresponding to the edge $(i,j)$.
Furthermore, since the block-graph $\mathcal{G}_b$ is sparse, an augmenting path
can be found fairly quickly. 

Furthermore, similar to the BK algorithm, we find
an augmenting path $P_b$ using BFS and maintain the search tree throughout
the algorithm, by repairing it whenever the block-graph is updated.
However, since the block-graph needs to be reconstructed after each
augmentation, for simplicity, we maintain a
single tree\footnote{The BK algorithm maintains two trees, source-tree and
sink-tree, but we only maintain the source-tree.}.
More specifically, we grow the search tree from source (node 0), in a breadth
first manner, and if sink (node 1) is reached, then the augmenting path $P_b$ is
found.

\subsection{Augmentation in the Block-graph}\label{sec:aug}
Now, given an augmenting path $P_b$ in the block-graph, we want to push
the maximum permissible flow through it. 
More specifically, since $P_b$ corresponds to a set of augmenting paths
$\{p_b\}$ in the Ishikawa graph, we will push the maximum flow through each
path $p_b$, until no such path exists.
This could be achieved by constructing the subgraph
$\hcalG^p$ of the Ishikawa graph corresponding to the augmenting path
$P_b$, and then finding each of the augmenting path $p_b$ by searching in
$\hcalG^p$. This would require us to either store $\hcalG^p$ (not memory
efficient) or call the flow reconstruction algorithm too many times.

Instead, we propose breaking down the augmentation operation in the
block-graph into a sequence of flow-loops and a subtraction
along a column. Then, the maximum flow through the path can be pushed in a greedy manner, by
pushing the maximum flow through each flow-loop. Before
describing this procedure in detail, we introduce the following definitions.

\begin{dfn}
\label{dfn:loop}
A flow-loop $m(\lambda,\mu,\alpha)$ in the Ishikawa graph is defined as the 
following sequence of operations: First, a value $\alpha$ is pushed down 
the left column from $U_{i:\ell}$
to $U_{i:\lambda}$, then across from $U_{i:\lambda}$ to $U_{j:\mu}$, and finally
up the right column from $U_{j:\mu}$ to $U_{j:\ell}$. Thus, applying the
flow-loop $m(\lambda,\mu,\alpha)$ corresponds to replacing $\phi$ by $\phi+\Delta$,
where
\begin{align*}
\Delta_{i:\lambda'} &= -\alpha\quad \forall \lambda' \ge \lambda\ ,\\
\Delta_{ij:\lambda\mu} &= -\alpha\ ,\\
\Delta_{ji:\mu\lambda} &= \alpha\ ,\\
\Delta_{j:\mu'} &= \alpha\quad \forall \mu' \ge \mu\ .
\end{align*}
\end{dfn}

\begin{dfn}
A flow-loop $\tilde{m}(\gamma,\delta,\alpha)$ in the block-graph 
$\mathcal{G}_b$ is defined by the following sequence of operations: 
First a value $\alpha$ is pushed down the left column from $U_{i:\ell}$ to
$B_{i:\gamma}$, then across from $B_{i:\gamma}$ to $B_{j:\delta}$, and finally
up the right column from $B_{j:\delta}$ to $U_{j:\ell}$. 
\end{dfn}

\begin{figure}
\begin{center}
\includegraphics[width=0.85\linewidth, trim=1.8cm 4.0cm 5.2cm 0.7cm, clip=true,
page=69]{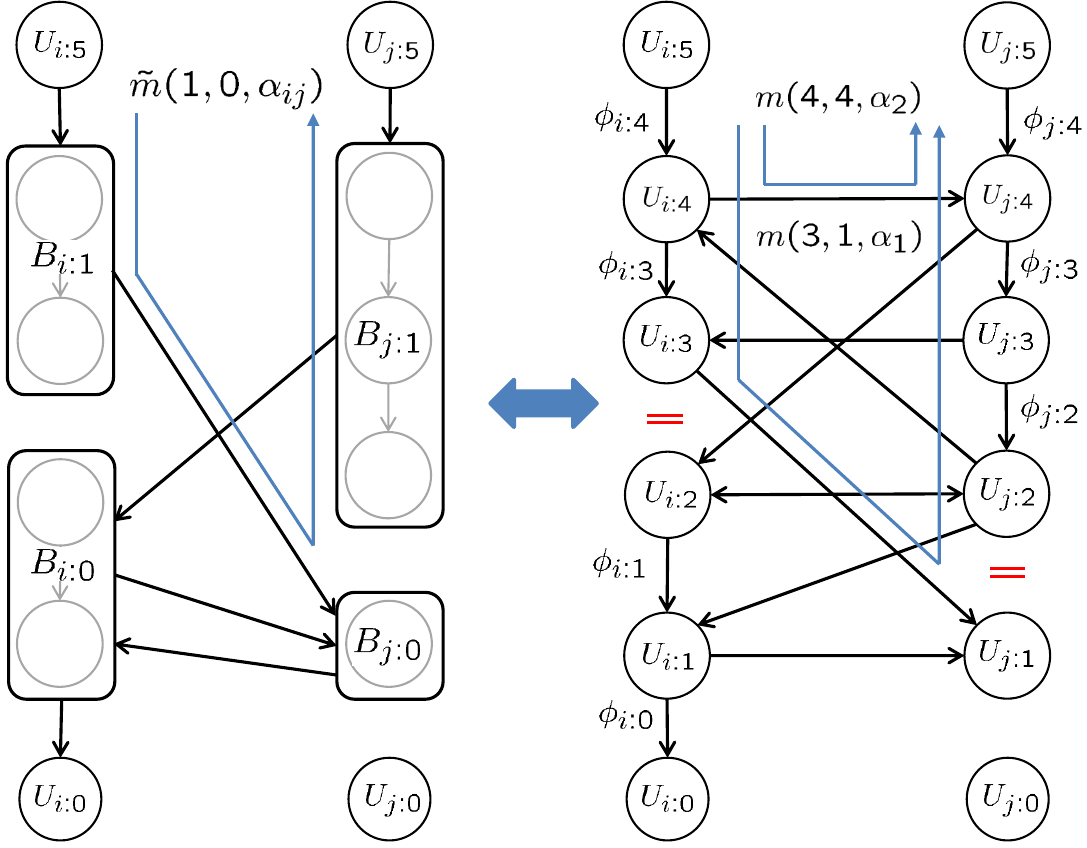}	
\end{center}
\vspace{-3ex}
\caption{\em An example flow-loop $\tilde{m}(1,0,\alpha_{ij})$ in the
block-graph (left) is equivalent to the summation of two flow-loops
$m(3,1,\alpha_1)$ and $m(4,4,\alpha_2)$ in the Ishikawa graph (right), with
$\alpha_{ij} = \alpha_1 + \alpha_2$.}
\label{fig:loop}
\vspace{-2ex}
\end{figure}

Note that, for a flow-loop $\tilde{m}(\gamma,\delta,\alpha)$ to be
permissible, block $B_{i:\gamma}$ must contain node $U_{i:\ell-1}$.
Note also that the flow-loop
$\tilde{m}(\gamma,\delta,\alpha)$ can be thought of as a summation of flow-loops
$m(\lambda,\mu,\alpha')$, where $U_{i:\lambda} \in B_{i:\gamma}$ and
$U_{j:\mu}\in B_{j:\delta'}$, for all $\delta' \ge \delta$ (see Fig.
\ref{fig:loop}). 

\begin{figure}
\begin{center}
\includegraphics[width=0.80\linewidth, trim=1.9cm 7.1cm 7.3cm 1.3cm, clip=true,
page=77]{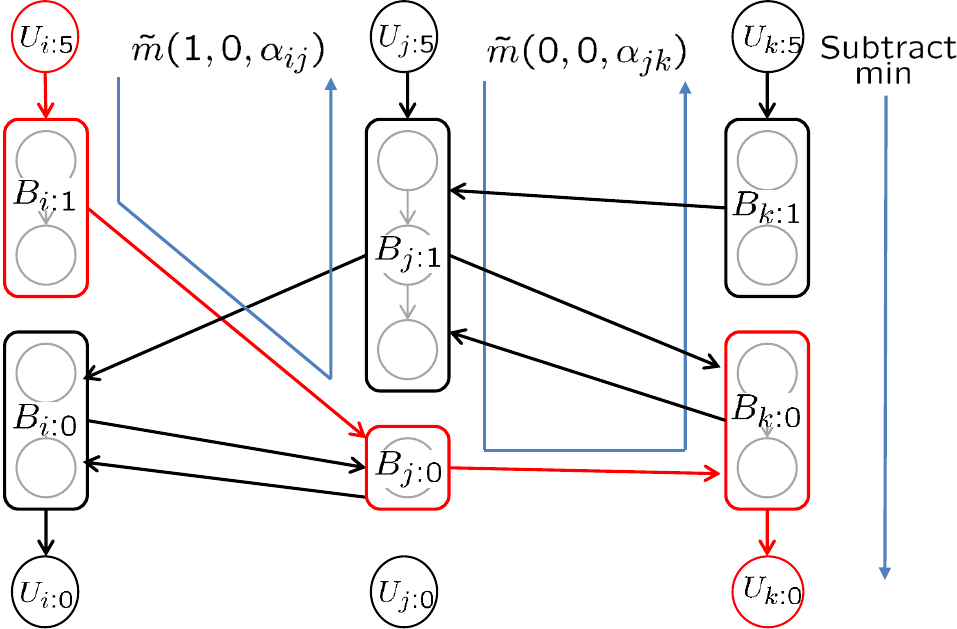}	
\end{center}
\vspace{-3ex}
\caption{\em An augmentation operation is broken down into a sequence of
flow-loops $\tilde{m}(\gamma,\delta,\alpha)$, and a subtraction along the 
column $k$. The augmenting path $P_s$ is highlighted in red.}
\label{fig:raug}
\vspace{-1ex}
\end{figure}

Given these definitions, one can easily see that the augmentation operation along the path
$P_b$ can be broken down into a sequence of flow-loops $\tilde{m}(\gamma,\delta,\alpha)$ 
and a subtraction along the last column $k$, as illustrated in Fig. \ref{fig:raug}. 
Now, we push the maximum permissible flow through $P_b$, using the
following greedy approach.

%
%
%

For each edge
$B_{i:\gamma}\to B_{j:\delta}$ that is part of the path $P_b$, we apply a 
flow-loop $\tilde{m}(\gamma,\delta,\alpha_{ij})$, where
$\alpha_{ij}$ is the maximum permissible flow through the edge
$B_{i:\gamma}\to B_{j:\delta}$. 
In fact, applying this flow-loop translates to reconstructing the Ishikawa edge capacities
$\phi_{ij}$ corresponding to edge $(i,j)$ and then applying flow-loops $m(\lambda,\mu,\alpha')$
for all $\lambda \ge \check{\lambda}$ and $\mu \ge \check{\mu}$,
starting from $\check{\lambda}$ and $\check{\mu}$, until no permissible
flow-loop $m(\lambda,\mu,\alpha')$ exists, with $\check{\lambda}$ and
$\check{\mu}$ the smallest values such that $U_{i:\lambda} \in B_{i:\gamma}$
and $U_{j:\mu}\in B_{j:\delta}$.
%
Finally, in the last column $k$, all the values $\phi_{k:\lambda}$ are positive, 
and the minimum along column $k$ is
subtracted from each $\phi_{k:\lambda}$. It is easy to see that this approach
pushes the maximum permissible flow through the path $P_b$.

%

Since, for each edge $(i,j)$, we do not store all the $2\,\ell^2$ capacities, 
but only the $2\,\ell$ exit-flows $\Sigma$, augmentation must then
also update these values. Fortunately, there is a direct relation between the 
flow-loops and $\Sigma$. To see this, let us consider the example flow-loop 
$\tilde{m}(1,0,\alpha_{ij})$ shown in Fig.~\ref{fig:loop}. Applying this 
flow-loop updates the corresponding exit-flows
as 
\vspace{-0.2cm}
\begin{align}
\Sigma_{ij:3} = \Sigma_{ij:3} + \alpha_1\ ,\\\nonumber
\Sigma_{ji:1} = \Sigma_{ji:1} - \alpha_1\ ,\\\nonumber
\Sigma_{ij:4} = \Sigma_{ij:4} + \alpha_2\ ,\\\nonumber
\Sigma_{ji:4} = \Sigma_{ji:4} - \alpha_2\ .
\end{align}
Similar updates can be done for all flow-loops in our procedure.
Note that the edge $B_{i:\gamma} \to B_{j:\delta}$ represents a
collection of possible paths from all the nodes $U_{i:\lambda} \in B_{i:\gamma}$
to all the nodes $U_{j:\mu}\in B_{j:\delta'}$, for
all $\delta' \ge \delta$. Therefore, unlike in the full Ishikawa graph, after applying a
flow-loop, the portion of the graph $\mathcal{G}_b^{ij}$ corresponding to 
edge $(i,j)\in\mathcal{E}$ needs to be reconstructed. This, however, can be done 
in a memory efficient manner, since it only involves one edge $(i,j)$ at a time.

\begin{algorithm}[t]
\caption{Memory Efficient Max Flow (MEMF) - Efficient Version}
\label{alg:memf}
\begin{algorithmic}

\Require $\phi^0$
\Comment {Initial Ishikawa capacities} 

\State $\Sigma \gets 0, T \gets \emptyset$ 
\Comment{Initialize exit-flows and search tree}

\State $\calG_b \gets $ block-graph($\phi^0$)
\Comment{Initial block-graph}

\Repeat 

\State $(T, P_b) \gets $ augmenting\_path($\mathcal{G}_b, T$)
\Comment{Sec.~\ref{sec:find}}

\State $\Sigma \gets $ augment($P_b, \phi^0, \Sigma$)
\Comment{Sec.~\ref{sec:aug}}

\ForAll{edge $(i,j)\in \mathcal{E}$ affected by augmentation}

\State $\phi_{ij} \gets $ compute\_edges($\phi^0, \Sigma, i, j$)
\Comment{Sec.~\ref{sec:frec}}

\State $\calG_b^{ij} \gets $ block-graph($\phi_{ij}, i, j$)
\Comment{Sec.~\ref{sec:find}}

\EndFor

\State $T \gets $ repair\_tree($T, \calG_b$)
\Comment{Repair search tree}

\Until no augmenting paths possible 

\\\Return get\_labelling($T$)
\Comment{Read from search tree}

\end{algorithmic}
\end{algorithm}

\vspace{-0.3cm}
\subsection{Summary}\label{sec:sum}
Our memory efficient max-flow ({\bf MEMF}) method is summarized in Algorithm \ref{alg:memf}. 
Let us briefly explain the subroutines below.

\paragraph*{\textbf{block-graph}}
Given the initial Ishikawa parameters $\phi^0$, this subroutine constructs the
block-graph by amalgamating nodes into blocks as described in
Section~\ref{sec:find}. If the input to the subroutine is the Ishikawa
capacities $\phi_{ij}$ corresponding to the edge $(i,j)\in \calE$, then it
constructs the block-graph portion $\calG_b^{ij}$.

\paragraph*{\textbf{augmenting\_path}} 
Given the block-graph $\calG_b$ and the search tree $T$, this subroutine
finds an augmenting path $P_b$ by growing the search tree, as discussed in
Section~\ref{sec:find}.

\paragraph*{\textbf{augment}}
Given the path $P_b$, this subroutine pushes the maximum permissible flow
through it by applying flow-loops $\tilde{m}(\gamma, \delta, \alpha)$ and then
subtracting the minimum from the last column, as discussed in
Section~\ref{sec:aug}.

\paragraph*{\textbf{compute\_edges}}
This is the same subroutine as in Algorithm~\ref{alg:pmemf}. (see
Sec.~\ref{sec:spmemf}).

\paragraph*{\textbf{repair\_tree}}
This subroutine is similar to the \textit{adoption} stage of the BK algorithm.
Given the reconstructed block-graph, the search tree $T$ is repaired by
checking for valid \textit{parent} for each \textit{orphan} node. See section
3.2.3 in~\cite{boykov2004experimental} for more detail.

\paragraph*{\textbf{get\_labelling}}
This subroutine directly reads the optimal labelling from the search tree $T$.


As discussed Section~\ref{sec:spmemf}, the space complexity of our algorithm is
$\mathcal{O}(|\mathcal{E}|\,\ell)$. For the rest of the paper, this efficient
version of the algorithm is referred to as MEMF.


\section{Equivalence with Message passing}\label{sec:repar}
In this section, we will give a more insightful interpretation of our max-flow
algorithm, by showing equivalence with the min-sum message passing algorithm. In
particular, 
first we will characterize the notion of an augmenting path in the message
passing context. Then, we will show that, 
the max-flow algorithm is, in spirit, equivalent to min-sum message passing, 
for multi-label submodular MRFs.
Finally, we observe the relationship between the set of exit-flows in our
algorithm and the set of messages in the message passing algorithm.
To this end, first let us define the multi-label
graph which will be used to explain the equivalence.

\subsection{The Mutli-label Graph}
\begin{figure}
\begin{center}
\includegraphics[width=0.6\linewidth, trim=1.9cm 8.0cm 15.2cm 2.0cm, clip=true,
page=80]{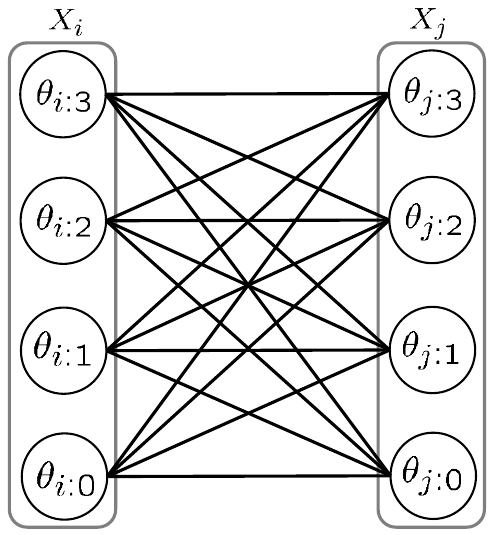}	
\end{center}
\vspace{-0.4cm}
\caption{\em Example of a multi-label grpah. Here the
nodes represent the unary potentials $\theta_{i:\lambda}$ and the edges
represent the pairwise potentials $\theta_{ij:\lambda\mu}$.}
\label{fig:mg}
\vspace{-0.5cm}
\end{figure}

An alternative way of representing the multi-label energy function (\ref{eqn:e})
is by defining \textit{indicator variables} $x_{i:\lambda} \in \{0,
1\}$, where $x_{i:\lambda} = 1$ if and only if $x_i = \lambda$. For a given $i$,
exactly one of $x_{i:\lambda}$\,;\, $\lambda \in \mathcal{L}$ can have value 1.
In terms of the indicator variables, the energy function (\ref{eqn:e}) may be written as
\begin{equation}
E_\theta(\mathbf{x}) = \sum_{i\in\mathcal{V}} \sum_{\lambda \in
\mathcal{L}} \theta_{i:\lambda}\, x_{i:\lambda} + \sum_{(i,j)\in\mathcal{E}} \sum_{\lambda,
\mu \in \mathcal{L}} \theta_{ij:\lambda\mu}\, x_{i:\lambda}\, x_{j:\mu}\ ,
\label{eqn:me}
\end{equation}
where the values $\theta$ are a particular set of parameters determining the
energy function.
One may define a graph, called a \textit{multi-label graph}, with nodes denoted 
by $X_{i:\lambda}\,;\, i \in \mathcal{V},\,  \lambda \in
\mathcal{L}$, as shown in Fig.~\ref{fig:mg}. This graph represents the energy
function. Given a labelling $\mathbf{x}$, the value of the energy function is
obtained by summing the weights on all nodes with $x_{i:\lambda} = 1$ (in other
words $x_i = \lambda$) plus the weights $\theta_{ij:\lambda\mu}$ such that
$x_{i:\lambda} = 1$ and $x_{j:\mu} = 1$.

Furthermore, given the Ishikawa edge capacities $\phi$ the parameters $\theta$
can be calculated using Eq.~\ref{eqn:thph0}. Note that, in this case,
$E_\theta(\mathbf{x}) = E_\phi(\mathbf{x})$ for all labellings $\mathbf{x}$.

\subsubsection{Reparametrization}
The energy function~\eqref{eqn:me} can be written in different ways
as a sum of unary and pairwise terms . In particular, there may exist a
different set of parameters $\theta'$ such that $E_\theta(\mathbf{x}) =
E_{\theta'}(\mathbf{x})$ for all $\mathbf{x}$, denoted as $E_\theta \equiv
E_{\theta'}$.
The
conditions under which $E_\theta \equiv E_{\theta'}$ are well known
\cite{kolmogorov2006convergent,werner2007linear}.

\begin{lem}
\label{lem:eqrep}
Two energy functions $E_\theta$ and $E_{\theta'}$ are equivalent if and only if
there exist values $m_{ij:\lambda}$ and $m_{ji:\mu}$ for $(i, j) \in
\mathcal{E}$ and $\lambda,\mu \in \mathcal{L}$ such that
\begin{align*}
\theta'_{ij:\lambda\mu} &= \theta_{ij:\lambda\mu} - m_{ij:\lambda} -
m_{ji:\mu}\ ,\\
\theta'_{i:\lambda} &= \theta_{i:\lambda} + \sum_{(k,i)\in\mathcal{E}^+}
m_{ik:\lambda}\ .
\end{align*}
\end{lem}

The values of $m_{ij:\lambda}$ constitute a \textit{message} $m_{ij}$ passed
from the edge $(i, j)$ to the node $i$; it may be thought of as a message
vector (indexed by $\lambda$).
A message $m_{ij}$ causes values $m_{ij:\lambda}$ to be
\textit{swept out} of all the edges $\theta_{ij:\lambda\mu}$ and adds onto
the nodes $\theta_{i:\lambda}$. Messages are passed in both directions from an
edge $(i, j)$. 

Note that, reparametrization provides an alternative way of
implementing the min-sum message passing algorithm. In particular, the objective
of min-sum message passing is to reparametrize the energy function so that the
constant term $\theta_c$ is maximized, while keeping the parameters $\theta$
non-negative, where the constant term is defined as follows:
\begin{equation}\label{eqn:tc}
\theta_c = \sum_{i\in \calV} \min_{\lambda\in \calL} \theta_{i:\lambda}\ . 
\end{equation}
For more details see~\cite{wainwright2005map,
kolmogorov2006convergent}.

\subsection{The Notion of an Augmenting Path in the Multi-label Graph}
We will now give an equivalent notion of an augmenting path in the context of
the multi-label graph. To this end, we will first understand the motivation
behind finding an augmenting path in the Ishikawa graph.

Let us consider an augmenting path in the Ishikawa
graph. If it is a \textit{trivial} augmenting path, \ie, it is an
augmenting path along a column from nodes $U_{i:\ell}$ to $U_{i:0}$, then
pushing the maximum flow along the path translates into subtracting
the minimum value from each $\phi_{i:\lambda}\,;\, \lambda \in \calL$.
In the multi-label graph, it is trivially equivalent to subtracting
the minimum from each
$\theta_{i:\lambda}\,;\, \lambda \in \calL$.

Let us consider a more interesting augmenting path in the Ishikawa graph, which
contains at least one cross edge $e_{ij:\lambda\mu}\in \hcalE_c$. Similar to the
discussion in Section~\ref{sec:aug}, one can easily see that, the augmentation operation can
be broken down into a sequence of flow-loops $m(\lambda, \mu, \alpha)$ (see
Def.~\ref{dfn:loop}) and a subtraction along a column. This intuitively suggests
that an augmenting path in the Ishikawa graph can be translated into a trivial augmenting 
path by passing flow around loops, \ie, they differ by a \textit{null} flow.
Therefore, the motivation of finding an augmenting path, is to pass flow around loops 
to get a trivial augmenting path.

Note that, the notion of a trivial augmenting path in
the multi-label graph is,
\begin{equation}\label{eqn:ta}
\theta_{i:\lambda} > 0\quad \forall\, \lambda \in \calL\ ,
\end{equation}
for some $i \in \calV$. Furthermore, by Lemma~\ref{lem:eqflow} and
by Lemma~\ref{lem:eqrep}, a flow-loop (or null flow) corresponds to a
reparametrization of the multi-label energy function. Hence, the notion of an
augmenting path in the multi-label graph can be characterized as, finding a set
of reparametrizations that makes $\theta_{i:\lambda}$ positive for all
$\lambda\in \calL$ for some $i\in \calV$.

\subsection{Equivalence of Max-flow and Min-sum Message Passing}
Note that, as we have discussed above, an augmenting path in the
Ishikawa graph can be translated into a trivial augmenting path by passing flow
around loops. Furthermore, pushing the maximum flow through a trivial augmenting
path is simply a subtraction of the minimum value $\min_{\lambda\in \calL}
\phi_{i:\lambda}$ from each $\phi_{i:\lambda}$. In fact, one can accumulate the
total flow passed from source to sink, which is exactly the constant term
defined in Eq.~\ref{eqn:tc}. Hence, max-flow tries to maximize $\theta_c$, by
passing flow around loops (\ie, reparametrizing), while keeping the edge
capacities $\phi$ non-negative. This is, in spirit, equivalent to the min-sum
message passing algorithm. Note that, the optimality of min-sum message passing
for the case of multi-label submodular MRFs,
is observed in~\cite{werner2007linear,kolmogorov2012optimality}.

\subsection{Flow-loop as a Reparametrization}
As mentioned above, from Lemma~\ref{lem:eqflow} and 
Lemma~\ref{lem:eqrep}, it is clear that, a flow-loop corresponds to a 
reparametrization of the multi-label energy function. In this section, we
will find the equivalent reparametrization of a flow-loop $m(\lambda, \mu,
\alpha)$. This will later allow us to understand the relationship between the
set of exit-flows and the set of messages. Let us now state and prove our
theorem.

\begin{thm}
\label{thm:fr}
Applying a flow-loop $m(\lambda,\mu,\alpha)$ in the Ishikawa graph 
is equivalent to a reparametrization in the multi-label graph, with messages
\begin{align}
m_{ij:\lambda'} &= -\alpha\quad \forall\, \lambda' \ge \lambda \ ,\\\nonumber
m_{ji:\mu'} &= \alpha\quad \forall\, \mu' \ge \mu \ .
\end{align}
\end{thm}
\begin{proof}
Let the Ishikawa parameters be $\phi$ and the multi-label graph parameters be
$\theta$ and assume that the flow is applied between columns $i$ and $j$.
Also after the flow the parameters be $\phi'$ and $\theta'$ respectively. 
Since $\theta$ can be calculated from $\phi$ using Eq.~\ref{eqn:thph0},
$E_\theta \equiv E_\phi$. Similarly $E_{\phi'} \equiv E_{\theta'}$. Also from
Lemma~\ref{lem:eqflow}, $E_{\phi} \equiv E_{\phi'}$. 
Hence, $E_\theta \equiv E_\phi \equiv E_{\phi'} \equiv E_{\theta'}$.
Now from Definition~\ref{dfn:loop},
\begin{align}
\phi'_{i:\lambda'} &= \phi_{i:\lambda'} - \alpha\quad \forall\, \lambda' \ge
\lambda\ ,\\\nonumber
\phi'_{j:\mu'} &= \phi_{j:\mu'} + \alpha\quad \forall\, \mu' \ge \mu\
,\\\nonumber
\phi'_{ij:\lambda\mu} &= \phi_{ij:\lambda\mu} - \alpha\ ,\\\nonumber
\phi'_{ji:\mu\lambda} &= \phi_{ji:\mu\lambda} + \alpha\ .
\end{align}
Substituting in Eq.~\ref{eqn:thph0},
\begin{align}
\theta'_{i:\lambda'} &= \theta_{i:\lambda'} - \alpha\quad \forall\, \lambda' \ge
\lambda\ ,\\\nonumber
\theta'_{j:\mu'} &= \theta_{j:\mu'} + \alpha\quad \forall\, \mu' \ge \mu\
,\\\nonumber
\theta'_{ij:\lambda'\mu'} &= \theta_{ij:\lambda'\mu'} - \alpha \quad \forall\,
\lambda' < \lambda,\ \mu' \ge \mu\ ,\\\nonumber
\theta'_{ij:\lambda'\mu'} &= \theta_{ij:\lambda'\mu'} + \alpha \quad \forall\,
\lambda' \ge \lambda,\ \mu' < \mu\ .
\end{align}
Now, since $E_\theta \equiv E_{\theta'}$, by Lemma~\ref{lem:eqrep}, there exists
messages $m_{ij:\lambda}$ and $m_{ji:\mu}$ such that,
\begin{align}
\theta'_{ij:\lambda\mu} &= \theta_{ij:\lambda\mu} - m_{ij:\lambda} -
m_{ji:\mu}\ ,\\\nonumber
\theta'_{i:\lambda} &= \theta_{i:\lambda} + \sum_{(k,i)\in\mathcal{E}^+} m_{ki:\lambda}\ .
\end{align}
With a little bit of calculation, one can see that, the
messages take the following form
\begin{align}
m_{ij:\lambda'} &= -\alpha\quad \forall\, \lambda' \ge \lambda \ ,\\\nonumber
m_{ji:\mu'} &= \alpha\quad \forall\, \mu' \ge \mu \ .
\end{align}
Note that, for a permissible flow
$m(\lambda,\mu,\alpha)$, the parameters $\phi'$ and $\theta'$ are
non-negative. 
\end{proof}

\begin{figure}
\begin{center}
\includegraphics[width=\linewidth, trim=0.6cm 6.0cm 4.8cm 1.5cm,
clip=true, page=81]{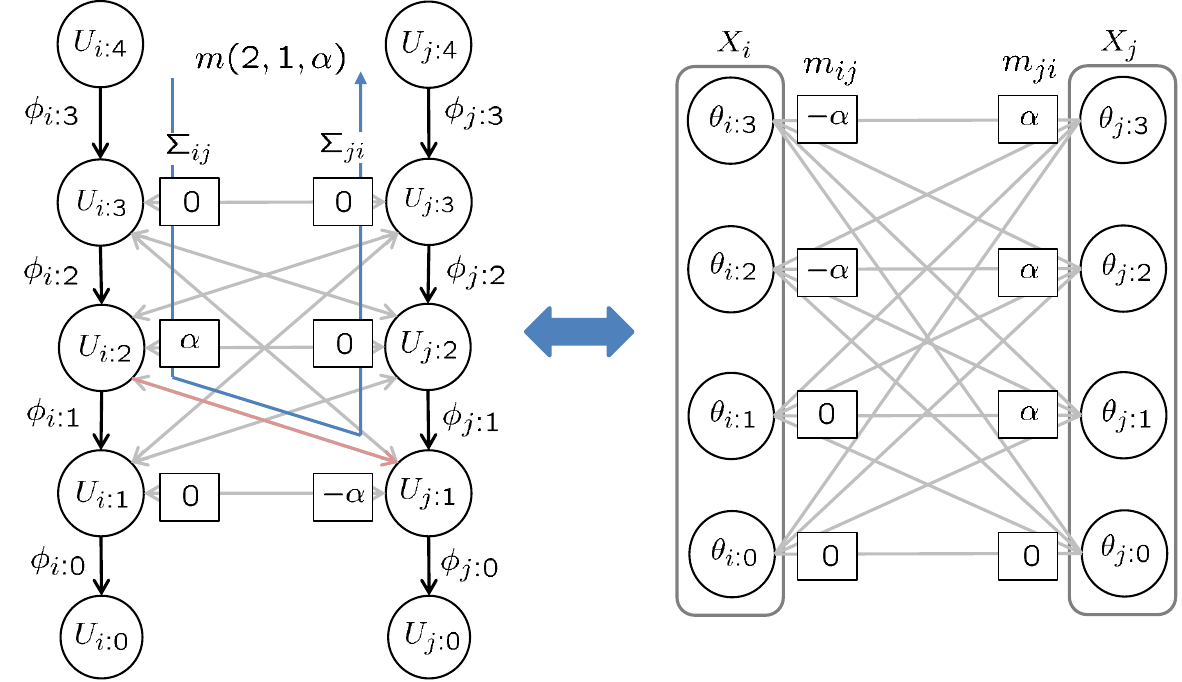}	
\end{center}
\vspace{-3ex}
\caption{\em A flow $m(2,1,\alpha)$ in the Ishikawa graph (left) and
its equivalent reparametrization in the multi-label graph (right). Note that,
the exit-flow vectors ($\Sigma_{ij}, \Sigma_{ji}$) and the corresponding message
vectors ($m_{ij}, m_{ji}$) are shown next to the nodes.}
\label{fig:fr}
\end{figure}

This equivalence is shown in Fig.~\ref{fig:fr} for an example flow-loop $m(2, 1,
\alpha)$. Note that, as shown in the figure, the flow $\alpha$ through
an edge $\phi_{ij:\lambda\mu}$ may be recorded in the set of exit-flows
$\Sigma$. Furthermore, as shown in the figure, the relationship between the
set of exit-flows and the set of messages can be written as,
\begin{align}
\label{eqn:fm}
\Sigma_{ij:\lambda} &= m_{ij:\lambda-1} - m_{ij:\lambda} \quad \forall\, \lambda
\in \{1, \cdots, \ell-1\} \ ,\\\nonumber
\Sigma_{ji:\mu} &= m_{ji:\mu-1} - m_{ji:\mu} \quad \forall\, \mu
\in \{1, \cdots, \ell-1\} \ .
\end{align}


\section{Related work}\label{sec:rel}
The approaches that have been proposed to minimize
multi-label submodular MRFs can be roughly grouped into two categories: Those based on max-flow and those based on an LP relaxation of the problem. Below, we briefly review representative techniques in each category.

\vspace{-0.25cm}
\subsection{Max-flow-based Methods} 
The most popular method to minimize a
multi-label submodular MRF energy is to construct the Ishikawa graph~\cite{ishikawa2003exact} and then apply a max-flow algorithm to find the
min-cut solution. Broadly speaking, there are three different kinds
of max-flow algorithms: those relying on finding augmenting paths~\cite{ford1962flows}, the push-relabel approach~\cite{goldberg1988new} and the pseudo-flow techniques~\cite{chandran2009computational}. Even though
numerous implementations are available, the BK
method~\cite{boykov2004experimental} is arguably the fastest implementation for 2D and sparse 3D graphs. Recently, for dense problems, the IBFS algorithm~\cite{goldberg2011maximum} was
shown to outperform the BK method in a number of
experiments~\cite{verma2012maxflow}. All the above-mentioned algorithms, however, require the same order of storage as the Ishikawa graph and hence scale poorly.
Two approaches have nonetheless been studied to scale the max-flow
algorithms. The first one explicitly relies on the N-D grid
structure of the problem at hand~\cite{delong2008scalable,jamrivska2012cache}. The second one makes use of distributed computing~\cite{shekhovtsov2013distributed,strandmark2010parallel,vineet2008cuda}.
Unfortunately, both these approaches require additional resources (disk space or clusters) to
run max-flow on an Ishikawa graph. By contrast, our algorithm lets us efficiently minimize the energy of much larger
Ishikawa-type graphs on a standard computer. Furthermore, using the method of~\cite{strandmark2010parallel}, it can also be parallelized.

\vspace{-0.25cm}
\subsection{LP Relaxation-based Methods} 
One memory-efficient way to minimize a multi-label submodular MRF energy consists of formulating the
problem as a linear program and then maximize the dual using
message-passing techniques~\cite{wainwright2005map}.
Many such algorithms have been studied~\cite{kolmogorov2006convergent,komodakis2011mrf,SavchynskyyUAI2012,werner2007linear}.
Even though these algorithms are good at approximating the
optimal solution (also theoretically optimal for multi-label submodular
MRFs~\cite{kolmogorov2012optimality}), as evidenced by the comparison
of~\cite{kappes-2015-ijcv} and by our experiments, they usually take much
longer to converge to the optimal solution than max-flow-based techniques.


\section{Experiments}\label{sec:expr}
\begin{figure}
\begin{center}
\includegraphics[width=0.49\linewidth]{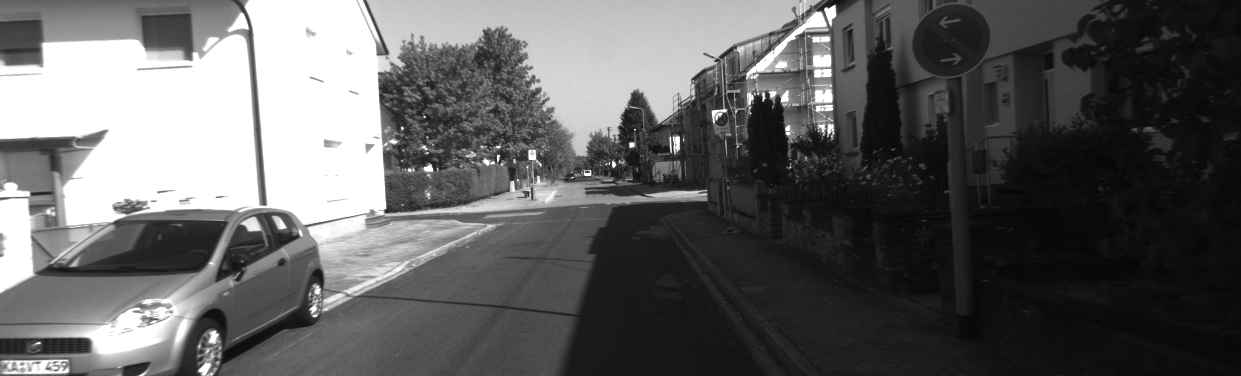}%
\hfill
\includegraphics[width=0.49\linewidth]{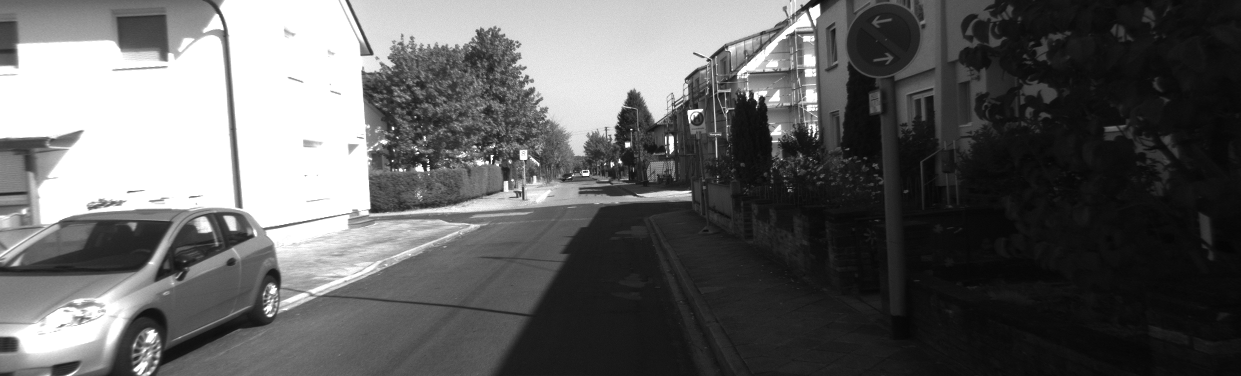}
\end{center}
\vspace{-3ex}
\caption{\em Left and right images of the stereo instance from
the KITTI dataset. The images are of size $1241\times376$, and we set the number of
labels to 40. This image pair was chosen arbitrarily as a representative of the dataset.}
\label{fig:kit}
\end{figure}

We evaluated our algorithm on the problems of stereo correspondence estimation
and image inpainting. For stereo correspondence estimation, we employed six
instances from the Middlebury dataset~\cite{scharstein2002taxonomy,scharstein2003high}: Tsukuba, Venus,
Sawtooth, Map, Cones and Teddy, and one instance from the KITTI dataset~\cite{geiger2013vision} (see Fig. \ref{fig:kit}). For Tsukuba and
Venus, we used the unary potentials of~\cite{szeliski2008comparative}, and for
all other stereo cases, those of~\cite{birchfield1998pixel}. 
For inpainting, we used the Penguin and House
images employed in~\cite{szeliski2008comparative}, and we used the same unary
potentials as in~\cite{szeliski2008comparative}. In all the above cases, 
we used pairwise potentials that can be expressed as
\begin{equation}
\theta_{ij}(x_i, x_j) = w_{ij}\, \theta(\left|x_i-x_j\right|)\;,
\end{equation} 
where, unless stated otherwise, the regularizer $\theta(\left|x_i-x_j\right|)$ 
is the quadratic function.
Furthermore, we employed a 4-connected neighbourhood structure, in all our
experiments.

We compare our results with two max-flow implementations:
the BK algorithm~\cite{boykov2004experimental} and Excesses
Incremental Breadth First Search (EIBFS)~\cite{goldberg2015faster} (which we
ran on the Ishikawa graph), and three LP relaxation-based algorithms:
Tree Reweighted Message Passing (TRWS)~\cite{kolmogorov2006convergent}, 
Subgradient based Dual Decomposition (DDSG)~\cite{komodakis2011mrf} 
and the Adaptive Diminishing Smoothing algorithm
(ADSal)~\cite{SavchynskyyUAI2012}.
For DDSG and ADSal, we used the Opengm~\cite{andres2012opengm} implementations. 
For the other algorithms, we employed the respective
authors' implementations. 

In practice, we only ran the BK algorithm and EIBFS if
the graph could be stored in RAM. Otherwise, we provide an estimate of their memory requirement. For LP relaxation-based methods, unless they converged, we ran the algorithms either for 10000 iterations, or for 50000 seconds, whichever occurred first. Note that the running
times reported for our algorithm include graph construction. All our experiments were conducted on
a 3.4 GHz i7-4770 CPU with 16 GB RAM.

The memory consumption and running times of the algorithms are provided in
Table~\ref{tab:mtquad}. Altogether, our algorithm lets us solve much larger problems 
than the BK algorithm and EIBFS, and is an order of magnitude faster than
state-of-the-art message-passing algorithms. 

\begin{table*}[t]
\begin{center}
\begin{tabular}
{>{\raggedright\arraybackslash}m{1.2cm}|>{\raggedleft\arraybackslash}m{1.2cm}
>{\raggedleft\arraybackslash}m{1.2cm} >{\raggedleft\arraybackslash}m{0.75cm}
>{\raggedleft\arraybackslash}m{0.75cm} >{\raggedleft\arraybackslash}m{0.8cm}
|>{\raggedleft\arraybackslash}m{0.9cm} |>{\raggedleft\arraybackslash}m{0.2cm}
>{\raggedleft\arraybackslash}m{0.6cm} >{\raggedleft\arraybackslash}m{1.15cm}
>{\raggedleft\arraybackslash}m{1.15cm} >{\raggedleft\arraybackslash}m{1.15cm}
|>{\raggedleft\arraybackslash}m{0.9cm}}
\multirow{2}{*}{Problem} & \multicolumn{6}{c|}{Memory [MB]} &
\multicolumn{6}{c}{Time [s]}\\
& BK & EIBFS & DDSG & ADSal & TRWS & MEMF & BK & EIBFS & DDSG & ADSal & TRWS &
MEMF\\ 
\hline

Tsukuba&3195&2495&258&252&287&\textbf{211}&14&\textbf{4}&$>$9083&$>$7065&198&28\\
Venus&7626&5907&424&418&638&\textbf{396}&35&\textbf{9}&$>$18156&1884&206&59\\
Sawtooth&7566&5860&415&415&633&\textbf{393}&31&\textbf{8}&$>$16238&10478&455&35\\
Map&6454&4946&\textbf{171}&208&494&219&57&\textbf{9}&$>$9495&$>$1679&187&36\\
Cones&*72303&*55063&\textbf{657}&939&5024&1200&-&-&$>$50000&$>$17866&1095&\textbf{364}\\
Teddy&*72303&*55063&\textbf{659}&939&5025&1200&-&-&$>$50000&$>$50000&6766&\textbf{2055}\\
KITTI&*88413&*67316&\textbf{1422}&1802&6416&2215&-&-&$>$50000&$>$50000&$>$45408&\textbf{18665}\\\hline
Penguin&*173893&*130728&236&1123&\textbf{215}&663&-&-&$>$50000&$>$50000&$>$50000&\textbf{6504}\\
House&*521853&*392315&689&2389&\textbf{643}&1986&-&-&$>$50000&$>$50000&$>$50000&\textbf{9001}\\

\end{tabular}
\end{center}
\vspace{-0.5cm}
\caption{\em Memory consumption and runtime comparison with state-of-the-art
baselines for quadratic regularizer (see para.~2 of Sec.~\ref{sec:expr},
for details on the algorithms).
A ``*'' indicates a memory estimate, and ``$>$'' indicates that the algorithm
did not converge to the optimum within the specified time. 
Note that our algorithm has a memory consumption $\mathcal{O}(\ell)$ times lower
than the max-flow-based methods and is an order of magnitude faster than message-passing algorithms. 
Compared to EIBFS, our algorithm is only 4 -- 7 times slower, but requires 12 -- 23 times less memory, which
makes it applicable to more realistic problems. In all stereo problems, TRWS
cached the pairwise potentials in an array for faster retrieval, but in the case
of inpainting, it was not possible due to excessive memory requirement.}
\label{tab:mtquad}
\vspace{-2ex}
\end{table*}

\begin{figure}
\begin{center}
\includegraphics[width=\linewidth]{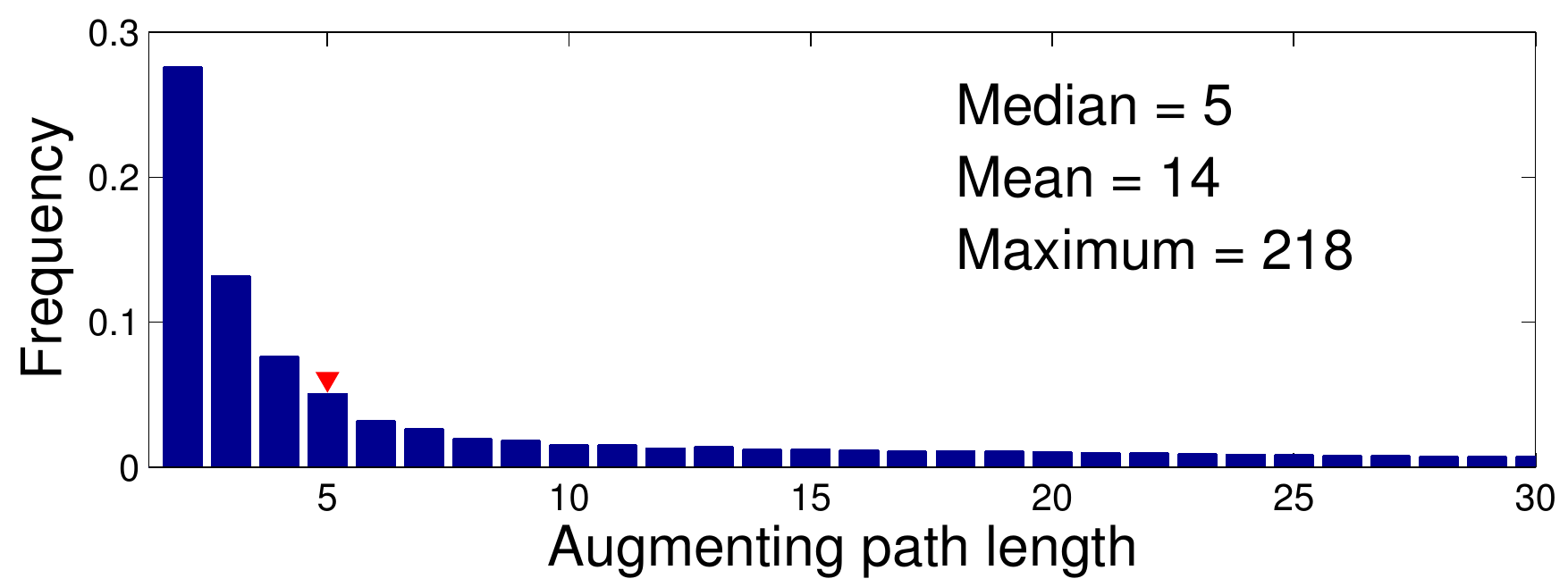}
\end{center}
\vspace{-0.5cm}
\caption{\em Lengths of augmenting paths found by our algorithm for
the Tsukuba stereo instance (see Sec.~\ref{sec:analy}). Each bar indicates the
proportion of paths of a certain length. For example, out of all augmenting paths $28\%$ of them were of length 2. The red arrow indicates the median length.}
\label{fig:pfreq}
\vspace{-2ex}
\end{figure}

\subsection{MEMF analysis}\label{sec:analy}
In this section, we empirically analyze various properties of our algorithm. 
First, note that, at each iteration, \ie, at
each augmentation step, our algorithm performs more computation than
standard max-flow. Therefore, we would like our algorithm to find short 
augmenting paths and to converge in fewer iterations than standard max-flow.
Below, we analyze these two properties empirically.

In Fig.~\ref{fig:pfreq}, we show the distribution of the lengths of the
augmenting paths found by our algorithm for the Tsukuba stereo instance. Note that the median length is only 5. 
As a matter of fact, the maximum length observed over all our experiments was 1073 for the KITTI data. 
Nevertheless, even in that image, the median length was only 15. Note that,
since our algorithm finds augmenting paths in the block-graph, the
path lengths are not directly comparable to those found by other
max-flow-based methods. In terms
of number of augmentations, we found that our algorithm only required between
35\% and 50\% of the total number of augmentations of the BK algorithm.

%
%
%

\begin{figure*}[t]
\begin{center}
\begin{subfigure}{.25\textwidth}
	\includegraphics[width=0.98\linewidth]{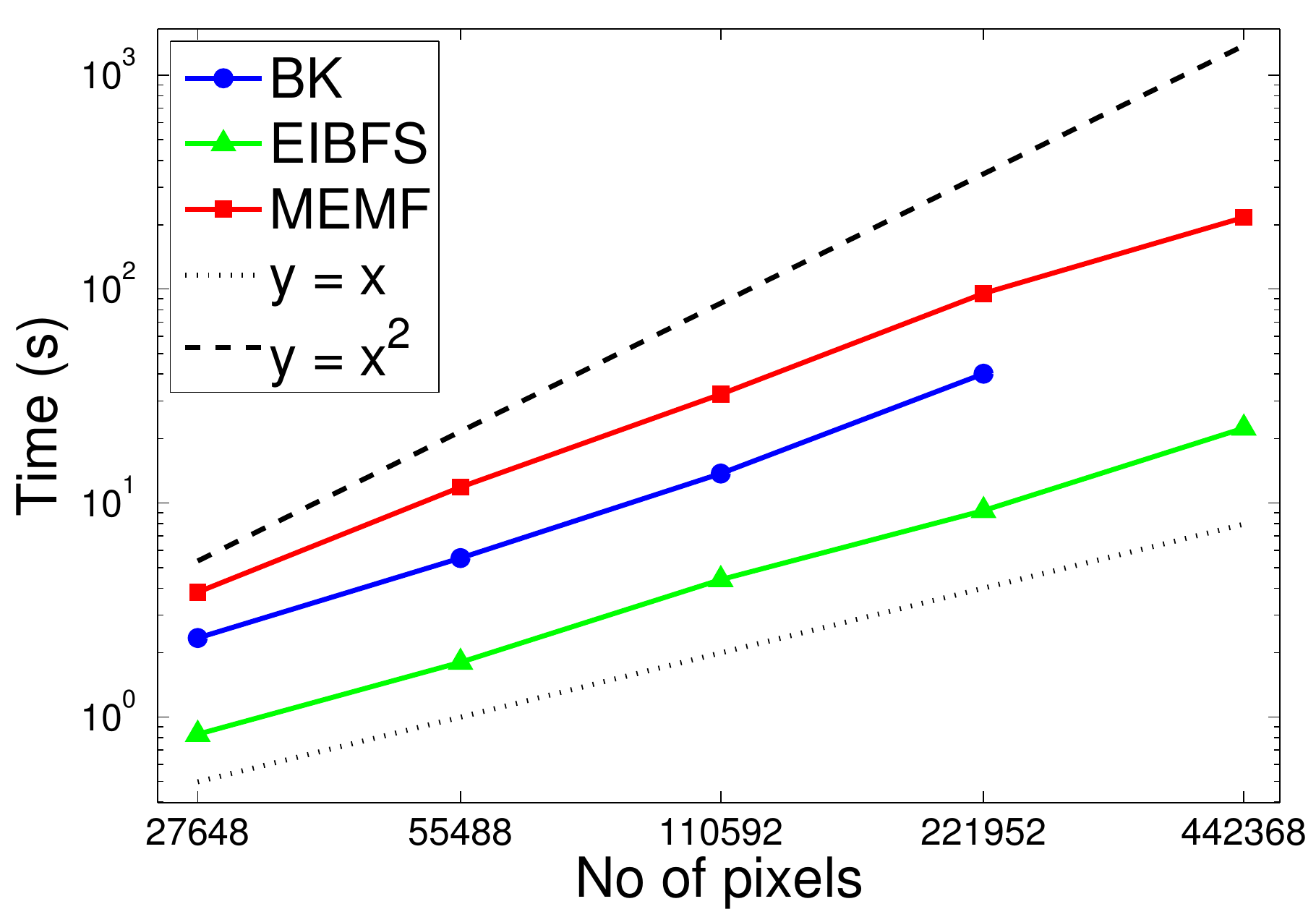}
	\caption{Tsukuba}
	\label{sfig:tms}
\end{subfigure}%
\begin{subfigure}{.25\textwidth}
	\includegraphics[width=0.98\linewidth]{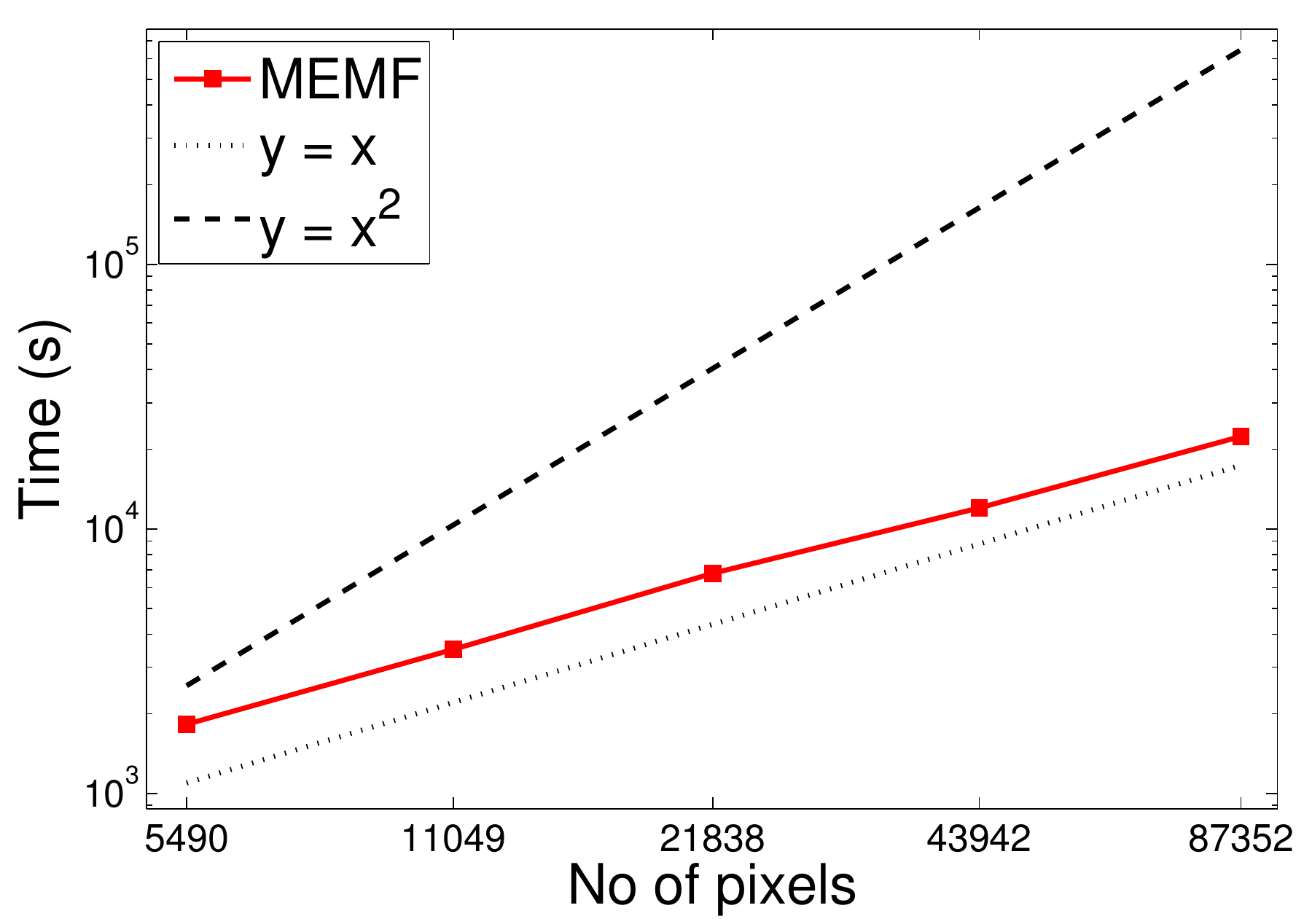}
	\caption{Penguin}
	\label{sfig:pms}
\end{subfigure}%
\begin{subfigure}{.25\textwidth}
	\includegraphics[width=0.98\linewidth]{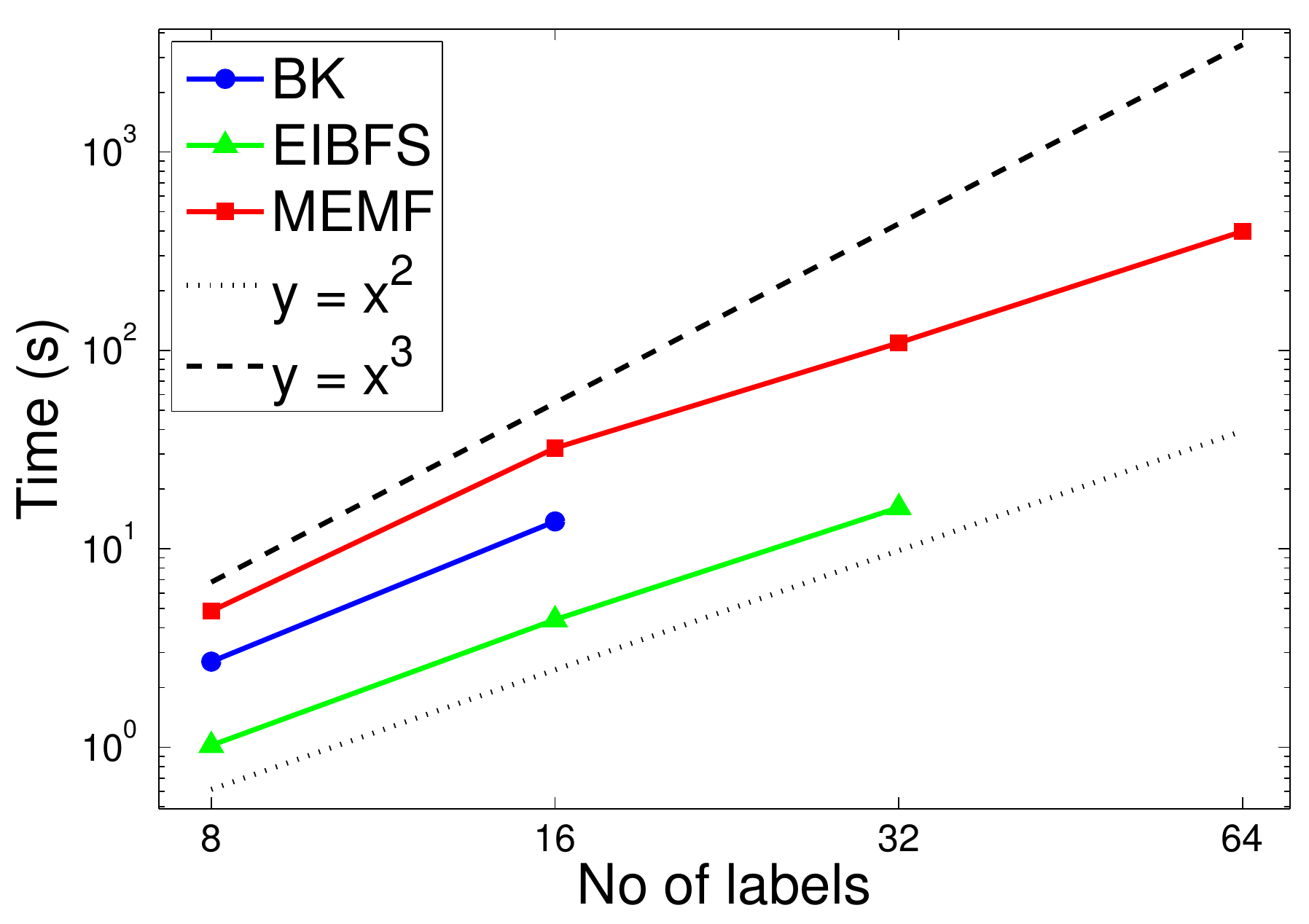}
	\caption{Tsukuba}
	\label{sfig:tml}
\end{subfigure}%
\begin{subfigure}{.25\textwidth}
	\includegraphics[width=0.98\linewidth]{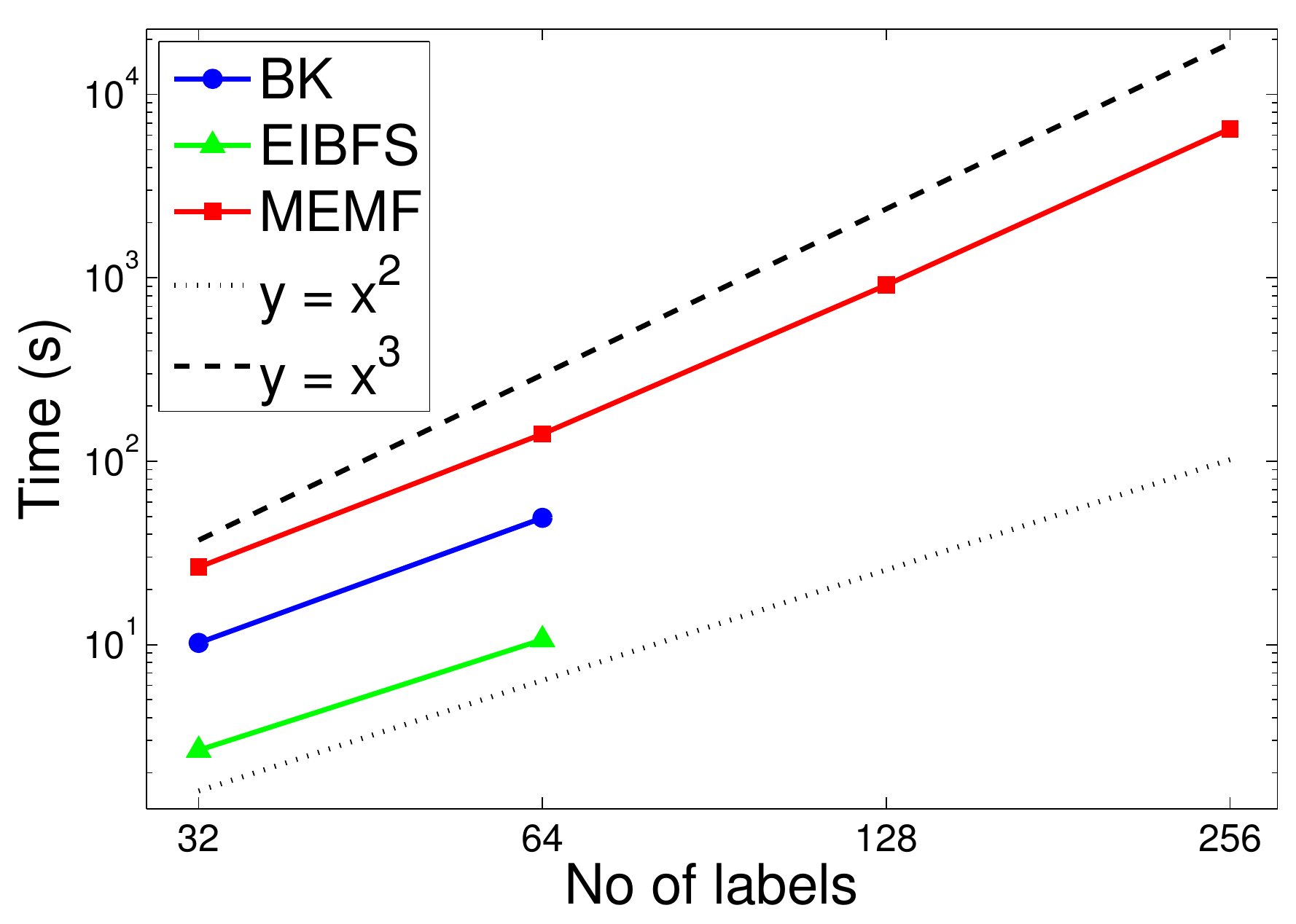}
	\caption{Penguin}
	\label{sfig:pml}
\end{subfigure}
\end{center}
\vspace{-0.4cm}
 \caption{\em Running time plots (in logarithmic scale) by changing the image
 size (\textbf{a}, \textbf{b}) and by changing the number of labels (\textbf{c},
 \textbf{d}), for Tsukuba and Penguin (see Sec.~\ref{sec:analy}).
 The dashed lines provide the reference slopes. Note that, all algorithms 
 exhibited near-linear performance with respect to the number of pixels and
 near-cubic performance with respect to the number of labels, but MEMF required
 $\calO(\ell)$ less memory.
 The plots of BK algorithm and EIBFS are not complete, since we could not 
 run them due to excessive memory requirement.}
\label{fig:msml}
\vspace{-0.2cm}
\end{figure*}

Next, we fixed the number of labels but varied the image size and compare the running
times of the max-flow algorithms, for Tsukuba and Penguin instances in
Fig.~\ref{sfig:tms},~\ref{sfig:pms}.
Similarly, we fixed the image size but varied the number of labels and report the running
times in Fig.~\ref{sfig:tml},~\ref{sfig:pml}. By doing this, we try to estimate
the empirical time complexity of our algorithm. Note that, similar to other
max-flow algorithms, MEMF exhibited near-linear performance with respect
to the image size and near-cubic performance with respect to the number of
labels, in these experiments. 

\begin{figure}
\begin{center}
\includegraphics[width=\linewidth]{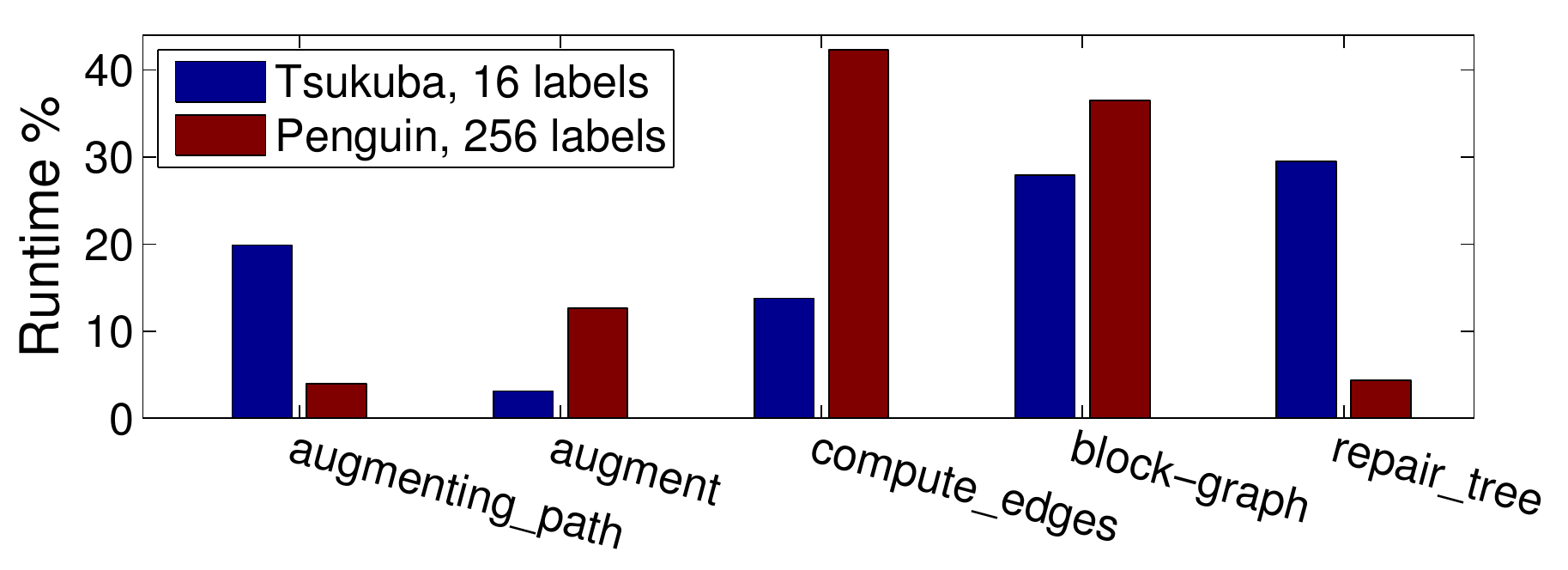}
\end{center}
\vspace{-0.6cm}
\caption{\em Percentage of time taken by each subroutine (see Sec.~\ref{sec:analy}).
Note that, in Penguin, due to large number of labels, the percentages of time spend on
\textit{compute\_edges} and \textit{block-graph} are high.}
\label{fig:st}
\vspace{-0.1cm}
\end{figure}

Finally, we report the percentage of time taken by each subroutine of our
algorithm, for Tsukuba and Penguin instances in Fig.~\ref{fig:st}. Note that
the individual time complexities of the subroutines \textit{compute\_edges} and
\textit{block-graph} are $\calO(\ell^3)$ and $\calO(\ell^2)$, respectively. 
Therefore, they become dominant when the number of labels is large, and hence the
corresponding percentages of time are high, particularly for Penguin.  

\begin{table}[t]
\begin{center}
\begin{tabular}
{>{\raggedright\arraybackslash}m{2.25cm}|>{\raggedleft\arraybackslash}m{1.15cm}
|>{\raggedleft\arraybackslash}m{1.0cm}|>{\raggedleft\arraybackslash}m{0.8cm}
|>{\raggedleft\arraybackslash}m{1.0cm}}
\multirow{2}{*}{Problem} &  \multicolumn{2}{c|}{Memory [MB]} &
\multicolumn{2}{c}{Time [s]}\\
 & BK & MEMF & BK & MEMF  \\
\hline

Penguin-128/10&4471&\textbf{332}&\textbf{224}&2566\\
House-64/15&8877&\textbf{498}&\textbf{106}&409\\
Penguin-256/20&*17143&\textbf{663}&-&\textbf{17748}\\
House-256/60&*137248&\textbf{1986}&-&\textbf{19681}\\

\end{tabular}
\end{center}
\vspace{-3ex}
\caption{\em Memory consumption and runtime comparison of IRGC+expansion
with either the BK method or our MEMF algorithm as
subroutine (see Sec.~\ref{sec:irgc}). Here, ``Penguin-128/10'' corresponds to
the Penguin problem with 128 labels and the truncated
quadratic function with truncation value 10 as regularizer. A ``*''
indicates a memory estimate. Compared to BK method, MEMF is only 4 -- 11 times
slower but requires 13 -- 18 times less memory, which makes it applicable to
much larger MRFs.}
\label{tab:irgcinp}
\vspace{-3ex}
\end{table}

\subsection{Minimizing Non-submodular MRFs}\label{sec:irgc}
Since our algorithm can simply replace standard max-flow in Ishikawa-type
graphs, we replaced the BK method with our MEMF procedure in the IRGC
algorithm~\cite{Ajanthan_2015_CVPR}, which minimizes MRFs with some non-convex
pairwise potentials (or regularizers) by iteratively building and solving an
Ishikawa graph.
This lets us tackle much larger non-submodular problems. In particular, we
computed inpainting results on Penguin by using all $256$ labels, as opposed to the down-sampled
label sets used in~\cite{Ajanthan_2015_CVPR}. The results of the IRGC+expansion
algorithm, with the BK method and with MEMF are
summarized in Table~\ref{tab:irgcinp}. 

\subsection{Robust Regularizer}\label{sec:hub}
Since robust regularizers are highly effective in computer vision, we tested our
algorithm by choosing Huber loss function~\cite{huber1964robust} as the
regularizer,
\begin{equation}
\theta(|x_i-x_j|) = \left\{ \begin{array}{ll} \frac{1}{2}|x_i-x_j|^2 & \mbox{if
$|x_i-x_j|\le \delta$}\\
\delta\left(|x_i-x_j|-\frac{1}{2}\delta\right) & \mbox{otherwise}\ ,\end{array}
\right.
\end{equation}
where $\delta$ is the Huber value.
The results are summarized in Table~\ref{tab:hub}.
In this experiment, the Huber value was set to 4 for Tsukuba, Venus and 
Sawtooth, 6 for Map, 20 for Cones and
Teddy, 10 for KITTI and 25 for Penguin and House.
Note that, the Ishikawa graph for a Huber regularizer is significantly
smaller, \ie, the number of edges per variable pair is $\calO(\delta\ell)$, instead of
$\calO(\ell^2)$.
Even in this case, our algorithm lets us solve much larger problems
than the BK algorithm and EIBFS, and is an order of magnitude faster than
state-of-the-art message-passing algorithms.

\begin{table*}[t]
\begin{center}
\begin{tabular}{l|rrr|r|rrr|r}
\multirow{2}{*}{Problem} & \multicolumn{4}{c|}{Memory [MB]} &
\multicolumn{4}{c}{Time [s]}\\
& BK & EIBFS & TRWS & MEMF & BK & EIBFS & TRWS &
MEMF\\ 
\hline

Tsukuba&1715&1385&287&\textbf{211}&8&\textbf{3}&198&28\\ 
Venus&3375&2719&638&\textbf{396}&17&\textbf{5}&211&57\\ 
Sawtooth&3348&2698&633&\textbf{393}&15&\textbf{4}&467&34\\ 
Map&2680&2116&494&\textbf{219}&22&\textbf{5}&$>$2953&36\\ 
Cones&*42155&*32167&5025&\textbf{1200}&-&-&1118&\textbf{363}\\ 
Teddy&*42155&*32167&5025&\textbf{1200}&-&-&6879&\textbf{2064}\\ 
KITTI&*42161&*32627&6416&\textbf{2215}&-&-&$>$30165&\textbf{18923}\\ \hline
Penguin&*33487&*25423&\textbf{215}&663&-&-&$>$50000&\textbf{6277}\\ 
House&*100494&*76295&\textbf{643}&1986&-&-&$>$50000&\textbf{8568}\\ 

\end{tabular}
\end{center}
\vspace{-0.4cm}
\caption{\em Memory consumption and runtime comparison with state-of-the-art
baselines for Huber regularizer (see Sec.~\ref{sec:hub}).
A ``*'' indicates a memory estimate, and ``$>$'' indicates that the algorithm
did not converge to the optimum within the specified time. 
Note that our algorithm has a much lower memory consumption than the max-flow-based methods and is an order of magnitude faster than message-passing algorithms. 
Compared to EIBFS, our algorithm is 7 -- 11 times slower, but requires 7 -- 10 times less memory, which
makes it applicable to more realistic problems. In all stereo problems, TRWS
cached the pairwise potentials in an array for faster retrieval, but in the case
of inpainting, it was not possible due to excessive memory requirement.}
\label{tab:hub}
\vspace{-0.5cm}
\end{table*}

\subsection{Parallelization}\label{sec:par}
\begin{figure}
\begin{center}
\includegraphics[width=\linewidth]{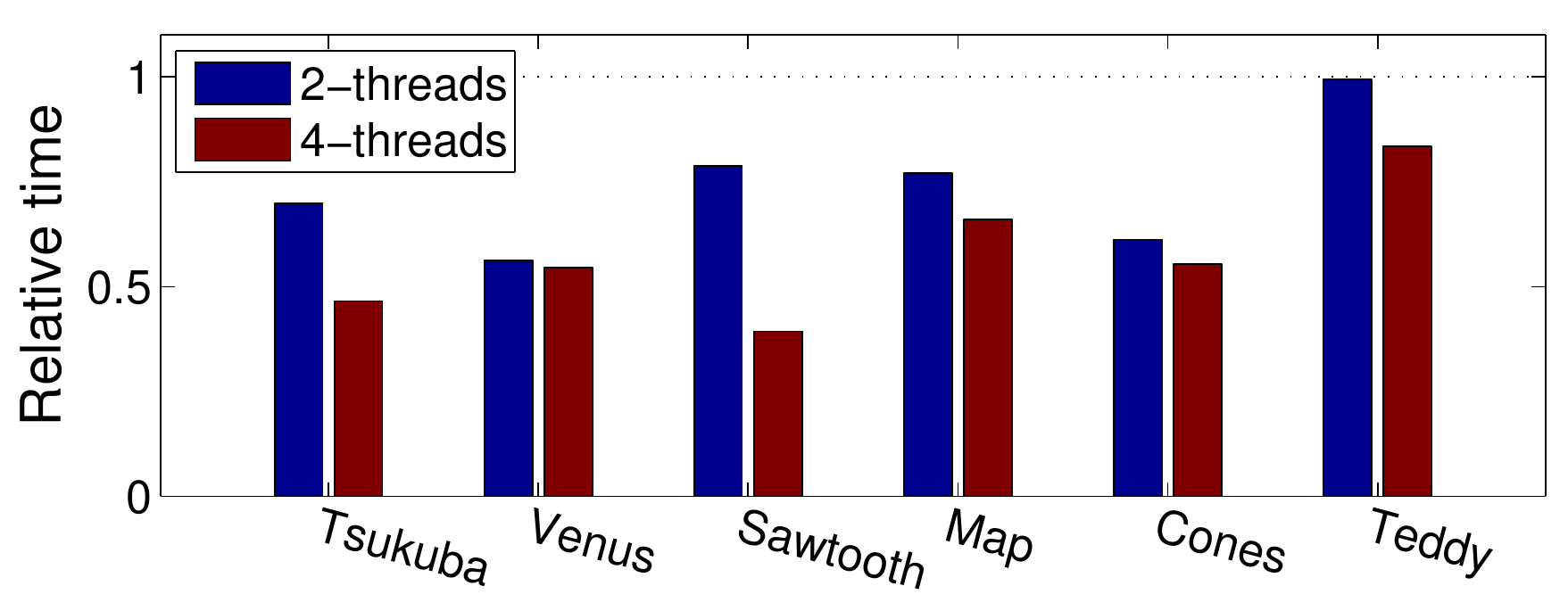}
\end{center}
\vspace{-0.6cm}
\caption{\em Our algorithm can be accelerated using the parallel max-flow
technique (see Sec.~\ref{sec:par}). The relative times ranged from 0.56 to 0.99
with 2-threads and from 0.39 to 0.83 with 4-threads. In Teddy, in the case of 2-threads, the
multi-threaded algorithm performs almost the same as the single-threaded
algorithm, which may be due to bad splits.}
%
\label{fig:pmemf}
\end{figure}

We parallelized our algorithm based on the dual-decomposition technique of~\cite{strandmark2010parallel} and evaluated on the same six stereo instances from the Middlebury dataset~\cite{scharstein2002taxonomy,scharstein2003high}. 
The relative times $t_m/t_s$, where $t_m$ stands for the multi-thread time and $t_s$ for the single-thread one, are shown in Fig.~\ref{fig:pmemf} for two and four threads. In this experiment, for all problems, the image grid was split vertically into two and four equally-sized blocks, respectively. Note that this spliting strategy is fairly arbitrary, and may affect the performance of the multi-threaded algorithm. In fact finding better splits may itself be a possible future direction.
%

\section{Conclusion}\label{sec:con}
We have introduced a variant of the max-flow algorithm that can minimize
multi-label submodular MRF energies optimally, while requiring much less storage.
Furthermore, our experiments have shown that our algorithm is an order of magnitude faster than 
state-of-the-art methods. We therefore believe that
our algorithm constitutes the method of choice to minimize Ishikwa type graphs when the complete 
graph cannot be stored in memory. 



%

\appendices

\section{Time Complexity Analysis of the Polynomial time
version of MEMF}\label{app:poly} Let us denote the Ishikawa graph with $\hcalG =
(\hcalV\cup\{0,1\}, \hcalE)$ and the lower-graph with $\hcalG^s =
(\hcalV\cup\{0,1\}, \hcalE^s)$\footnote{The superscript $s$ is used to restate
the fact, that the lower-graph is a subgraph of the Ishikawa graph.}.
In this section, we denote the nodes with $u$, $v$, etc.
Also the notation $u_1 \le u$ means, the node $u_1$ and $u$ are in the
same column where $u_1$ is below $u$. Let $\hcalG^s_f$ denotes the residual
graph of the lower-graph after the flow $f$ and similarly $\hcalE^s_f$ denotes
the set of non-zero residual edges. 
Let $d_f(u, v)$ denotes the shortest path distance from $u$ to $v$
calculated by MEMF.

\begin{lem}\label{lem:shp}
If the MEMF algorithm is run on the Ishikawa graph
$\hcalG = (\hcalV\cup\{0,1\}, \hcalE)$ with source $0$ and terminal $1$,
then for any node $v\in \hcalV$, the shortest path distance
$d_f(0, v)$ in the residual lower-graph $\hcalG^s_f$ increases monotonically
with each flow augmentation.
\end{lem}
\begin{proof}
We will suppose that for some node $v \in \hcalV$, there is a
flow augmentation that causes the shortest path distance from $0$ to $v$ to
decrease, and then we will derive a contradiction. Let $f$ be the flow just
before the first augmentation that decreases some shortest path distance, and
let $f'$ be the flow just afterward. Let $v$ be the node with the minimum
$d_{f'}(0, v)$ whose distance was decreased by the augmentation, so that
$d_{f'}(0, v) < d_f(0,v)$. Let $p = 0\leadsto u \to v$ be a shortest path from
$0$ to $v$ in $\hcalG^s_{f'}$, so that $(u, v)\in \hcalE^s_{f'}$ and
\begin{equation}
d_{f'}(0,v) = \left\{ \begin{array}{ll} d_{f'}(0, u) &
\mbox{if $u < v$ (infinite edge)}\\
d_{f'}(0, u) + 1 & \mbox{otherwise}\ .\end{array} \right.
\end{equation}
Because of how we chose $v$, we know that the distance of node $u$ from the
source $0$ did not decrease, \textit{i.e.},
\begin{equation}
d_{f'}(0,u) \ge d_f(0,u)\ . 
\end{equation}
We claim that $(u,v) \notin \hcalE^s_{f}$. Why? If we had $(u,v)\in
\hcalE^s_{f}$, then we would also have
\begin{align}\label{eqn:pr}
d_f(0,v) &\le d_f(0,u) + 1\ ,\\\nonumber
&\le d_{f'}(0,u) + 1\ ,\\\nonumber
&= d_{f'}(0,v)\ ,
\end{align}
which contradicts our assumption that $d_{f'}(0,v) < d_f(0,v)$. The above
argument simply follows even if $(u,v)$ is an infinite capacity edge. Hence
$(u,v) \notin \hcalE^s_{f}$.

How can we have $(u,v)\notin \hcalE^s_{f}$ and $(u,v)\in
\hcalE^s_{f'}$? Note that, in this case, $(u,v)$ cannot be an infinite capacity edge. There can
be two reasons:
  \begin{enumerate}
    \item A new lowest edge $(u,v)$ is created due to the flow from $v$ to $u$.
    That means the augmentation must have increased the flow from $v$ to $u$.
	The MEMF algorithm always augments flow along shortest paths, and therefore
	the shortest path from $0$ to $u$ in $\hcalG^s_f$ has $(v,u)$ as its last
	edge.
	Therefore,
	\begin{align}
	d_f(0,v) &= d_f(0, u) - 1\ ,\\\nonumber
	&\le d_{f'}(0, u) - 1\ ,\\\nonumber
	&= d_{f'}(0, v) - 2\ ,
	\end{align}
	which contradicts our assumption that $d_{f'}(0,v) < d_f(0,v)$.
    \item A new edge $(u,v)$ is created due to a saturating flow from $u$ to
    $v_1$ for some $v_1 < v$. The MEMF algorithm always augments flow along shortest paths, and
    therefore the shortest path from $0$ to $v_1$ in $\hcalG^s_f$ has
    $(u,v_1)$ as its last edge. Since $d_f(0,v)\le d_f(0,v_1)$, due to the
    upward infinite capacity edges, we have,
    \begin{align}
	d_f(0,v) &\le d_f(0, v_1)\ ,\\\nonumber
	&= d_f(0, u) + 1\ ,\\\nonumber
	&\le d_{f'}(0, u) + 1\ ,\\\nonumber
	&= d_{f'}(0, v)\ ,
	\end{align}
	which contradicts our assumption that $d_{f'}(0,v) < d_f(0,v)$.
  \end{enumerate}
 We conclude that our assumption that such a node $v$ exists is incorrect.
\end{proof}

The next theorem bounds the number of iterations of the MEMF algorithm.
\begin{thm}
If the MEMF algorithm is run on the Ishikawa graph
$\hcalG=(\hcalV\cup\{0,1\}, \hcalE)$ with source $0$ and sink $1$,
then the total number of augmentations performed by the algorithm is
$\mathcal{O}(|\hcalV||\hcalE|)$.
\end{thm}
\begin{proof}
We say that an edge $(u,v)$ in a residual lower-graph $\hcalG^s_f$
is \textit{critical} on an augmenting path $p$ if the residual capacity of $p$ is
the residual capacity of $(u,v)$, \textit{i.e.}, if $c_f(p) = c_f(u,v)$. After
we have augmented flow along an augmenting path, any critical edge on the path
disappears from the residual graph. Moreover, at least one edge on any
augmenting path must be critical. We will show that each of the
$|\hcalE|$ edges can become critical at most $|\hcalV|/2+1$ times.
Furthermore, note that, an infinite capacity edge cannot be critical at any point of the
algorithm.

Let $u$ and $v$ be nodes in $\hcalV\cup\{0,1\}$ that are connected by an edge in
$\hcalE^s$. Since augmenting paths are shortest paths, when $(u,v)$ is
critical for the first time, we have
\begin{equation}
d_f(0,v) = d_f(0,u) + 1\ .
\end{equation}
Once the flow is augmented, the edge $(u,v)$ disappears from the residual graph.
Since we maintain the lowest-cross-edge property, there cannot be an edge
$(u,v_1)$ in $\hcalG^s_f$ for some $v_1 < v$. Therefore, the edge $(u,v)$
cannot reappear later on another augmenting path until after the flow from $u$
to $v_1$ for some $v_1\le v$ is decreased, which occurs only if $(v_1,u)$
appears on an augmenting path. If $f'$ is the flow when 
this event occurs, then we have
\begin{equation}
d_{f'}(0,u) = d_{f'}(0,v_1) + 1\ .
\end{equation}
Since $d_{f'}(0,v)\le d_{f'}(0,v_1)$, due to the upward infinite capacity edges,
and $d_f(0,v) \le d_{f'}(0,v)$ by Lemma~\ref{lem:shp}, we have
\begin{align}
d_{f'}(0,u) &= d_{f'}(0,v_1) + 1\ ,\\\nonumber
&\ge d_{f'}(0,v) + 1\ ,\\\nonumber
&\ge d_{f}(0,v) + 1\ ,\\\nonumber
&= d_{f}(0,u) + 2\ .
\end{align}

Consequently, from the time $(u,v)$ becomes critical to the time when it next
becomes critical, the distance of $u$ from the source increases by at least 2.
The distance of $u$ from the source is initially at least 0. The intermediate
nodes on a shortest path from $0$ to $u$ cannot contain $0$, $u$ or $1$ (since
$(u,v)$ on an augmenting path implies that $u\ne 1$). Therefore, until $u$
becomes unreachable from the source, if ever, its distance is at most
$|\hcalV|$. Thus, after the first time that $(u,v)$ becomes critical,
it can become critical at most $|\hcalV|/2$
times more, for a total of $|\hcalV|/2+1$ times. Since there are
$\mathcal{O}(|\hcalE|)$ pairs of nodes that can have an edge between them
in a residual graph, the total number of critical edges during the entire
execution of the MEMF algorithm is
$\mathcal{O}(|\hcalV||\hcalE|)$. Each augmenting path has at least
one critical edge, and hence the theorem follows.
\end{proof}

\section{Proof of Theorem~\ref{thm:sim}}\label{app:sim}
\begin{thm1}
Given the set of Ishikawa graph parameters $\phi$, there is an augmenting path
in the block-graph if and only if there exists an augmenting
path in the Ishikawa graph.
\end{thm1}
\begin{proof}

First, we will prove that,
if there is an augmenting path in the block-graph, then there
exists an augmenting path in the Ishikawa graph.
It is clear that an augmenting path in the block-graph contains 
an edge from node 0 to a block and then a sequence of edges $B_{i:\gamma}\to B_{j:\delta}$ 
and finally an edge from a block to node 1. Note that an edge from node 0 to a 
block $B_{i:\gamma}$ corresponds to a positive edge $e_{i:\ell-1}$ in the 
Ishikawa graph; similarly an edge from a block $B_{j:\delta}$ to node 1 corresponds 
to a positive edge $e_{j:0}$.
Now, consider an edge $B_{i:\gamma}
\to B_{j:\delta}$ in the augmenting path. Corresponding to this, there exists a
positive edge $e_{ij:\lambda\mu}$ such that $U_{i:\lambda} \in B_{i:\gamma'}$
for some $\gamma' \ge \gamma$ and $U_{j:\mu}\in B_{j:\delta}$ in the Ishikawa graph. 
Also along
the column $i$, there are upward infinite capacity edges, and nodes corresponding 
to a block are also connected with positive bidirectional edges. Hence, there exists an augmenting
path in the Ishikawa graph, corresponding to the augmenting
path in the block-graph.

Now, we will prove the converse. Consider an augmenting path in the Ishikawa graph. 
The path may contain a sequence of positive edges
$e_{i:\lambda}$, $e_{ij:\lambda\mu}$ and infinite capacity edges
$e_{ii:\lambda\lambda+1}$. Note that, by construction, the $e_{i:\lambda}$
edges either will be in the same block $B_{i:\gamma}$ in the block-graph,
or will be between a block and node 0 or node 1.
Furthermore, the infinite capacity edges either will
be in the same block, or there will be an edge $B_{i:\gamma}\to B_{j:\delta}$ in the 
block-graph to represent them.
Finally, if $e_{ij:\lambda\mu}$ is a positive edge, then, by construction of the
block-graph, there exists an edge $B_{i:\gamma} \to B_{j:\delta'}$ where
$U_{i:\lambda}\in B_{i:\gamma}$ and $U_{j:\mu}\in B_{j:\delta}$ with $\delta'\le \delta$.
Hence,
if there is an augmenting path in the Ishikawa graph, then
there exists an augmenting path in the block-graph.
\end{proof}


%
%

\ifCLASSOPTIONcaptionsoff
  \newpage
\fi



\bibliographystyle{IEEEtran}
\bibliography{IEEEabrv,memf}
%
%
%

%

\begin{IEEEbiography}[{\includegraphics[width=1in,height=1.25in,clip,keepaspectratio]{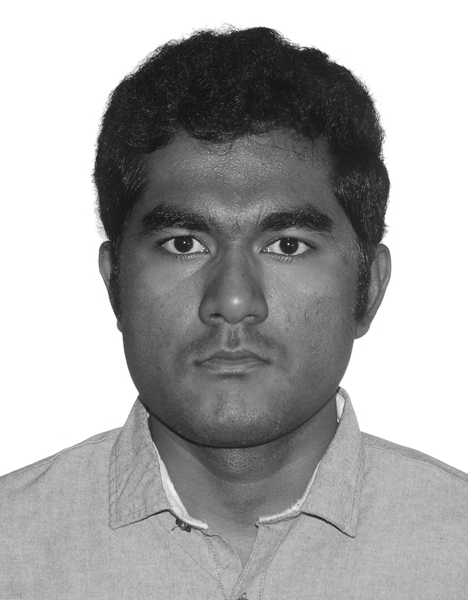}}]{Thalaiyasingam
Ajanthan} obtained his BSc degree
in Electronics and Telecommunication Engineering
from University of Moratuwa, Sri Lanka, in
2013. He is now a PhD student at the College of
Engineering and Computer Science, Australian
National University (ANU). He is also a member
of the Analytics Group at
Data61, CSIRO. He is a student member of IEEE.
\end{IEEEbiography}

\begin{IEEEbiography}[{\includegraphics[width=1in,height=1.25in,clip,keepaspectratio]{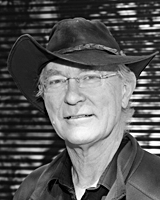}}]{Richard
Hartley} is a member of the Computer
Vision Group in the Research School of Engineering,
at ANU, where he has been since January,
2001. He is also a member of the Analytics group in Data61, CSIRO. He worked at
the GE Research and Development Center from
1985 to 2001, working first in VLSI design, and
later in computer vision. He became involved
with Image Understanding and Scene Reconstruction
working with GE’s Simulation and Control
Systems Division. He is an author (with A.
Zisserman) of the book Multiple View Geometry in Computer Vision.
\end{IEEEbiography}

\begin{IEEEbiography}[{\includegraphics[width=1in,height=1.25in,clip,keepaspectratio]{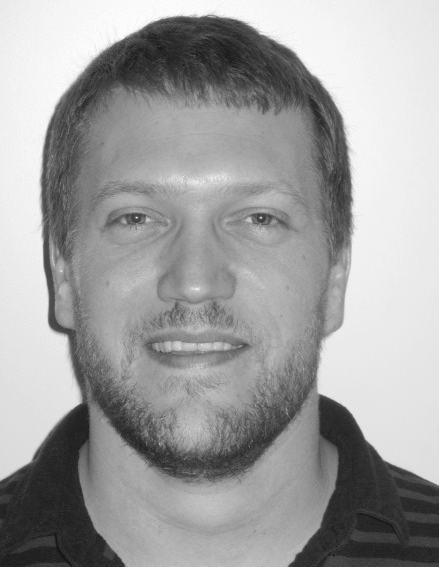}}]{Mathieu
Salzmann} obtained his MSc and PhD
degrees from EPFL in 2004 and 2009, respectively.
He then joined the International Computer
Science Institute and the EECS Department at
the University of California at Berkeley as a postdoctoral
fellow, later the Toyota Technical
Institute at Chicago as a research assistant professor and 
 a senior researcher at NICTA in Canberra. He is now a senior researcher in
 Computer Vision Lab at EPFL.
\end{IEEEbiography}







\end{document}